\documentclass[a4paper, 11pt, oneside]{article}
\pdfoutput=1
\usepackage{float}
\usepackage{jheppub}
\usepackage{epstopdf}
\usepackage{mathtools}
\usepackage{graphics}
\usepackage{multirow}
\newcommand{\nc}{\newcommand}

\newlength{\absize}
\nc{\non}{\nonumber}
\nc{\hc}{\hbox {H.c.}}
\nc{\noi}{\noindent}
\nc{\barx}{\bar{x}}
\nc{\pbarn}{\;\hbox {pb}}
\nc{\fbarn}{\;\hbox {fb}}
\nc{\lsp}{\;\;\;\;\;}
\nc{\Lsp}{\;\;\;\;\;\;\;\;\;\;}
\nc{\LLsp}{\lspace \lspace}
\nc{\lra}{\longrightarrow}
\nc{\beq}{\begin{equation}}  \nc{\eeq}{\end{equation}}
\nc{\bea}{\begin{eqnarray}}  \nc{\eea}{\end{eqnarray}}
\nc{\baa}{\begin{array}}     \nc{\eaa}{\end{array}}
\nc{\bit}{\begin{itemize}}   \nc{\eit}{\end{itemize}}
\nc{\ben}{\begin{enumerate}} \nc{\een}{\end{enumerate}}
\nc{\bce}{\begin{center}}    \nc{\ece}{\end{center}}
\nc{\bpm}{\begin{pmatrix}}   \nc{\epm}{\end{pmatrix}}
\nc{\bvt}{\begin{verbatim}}  \nc{\evt}{\end{verbatim}}

\newcommand{\textoverline}[1]{$\overline{\mbox{#1}}$}

\def\lsim{\mathrel{\raise.3ex\hbox{$<$\kern-.75em\lower1ex\hbox{$\sim$}}}}
\def\gsim{\mathrel{\raise.3ex\hbox{$>$\kern-.75em\lower1ex\hbox{$\sim$}}}}

\def\bcal{{\cal B}}

\def\lcal{{\cal L}}
\def\mcal{{\cal M}}

\def\pcal{{\cal P}}
\def\qcal{{\cal Q}}

\nc{\tanb}{\tan\beta}
\nc{\mch}{M_{H^\pm}}

\def\thetaW{{\theta}_{\rm W}}

\def\mch{M_{H^\pm}}
\def\mchsq{M_{H^\pm}^2}
\def\z2{Z_2}
\def\u1{U(1)}
\def\cp1{CP1}

\def\iff{{\it iff} }

\nc{\for}{\lsp {\rm for} \lsp}
\nc{\andd}{\lsp {\rm and} \lsp}

\renewcommand{\Re}{\mbox{Re\thinspace}}
\renewcommand{\Im}{\mbox{Im\thinspace}}
\newcommand{\half}{{\textstyle\frac{1}{2}}}

\allowdisplaybreaks 

\renewcommand{\vec}[1]{\boldsymbol{#1}}

\title{Softly broken symmetries in the 2HDM \\ --- an invariant formulation}
\author[a,b]{P.M. Ferreira,}
\affiliation[a]{Instituto Superior de Engenharia de Lisboa~---~ISEL, 1959-007~Lisboa, Portugal}
\affiliation[b]{Centro de F\'isica Te\'orica e Computacional, Faculdade de Ci\^encias,
Universidade de Lisboa, Campo Grande, 1749-016~Lisboa, Portugal}
\author[c]{B. Grzadkowski,}
\affiliation[c]{Faculty of Physics, University of Warsaw, Pasteura 5, 02-093 Warsaw, Poland}
\author[d]{O. M. Ogreid,}
\affiliation[d]{Western Norway University of Applied Sciences,
Postboks 7030, N-5020 Bergen, Norway, }
\author[e]{P. Osland}
\affiliation[e]{Department of Physics and Technology,
University of Bergen, Postboks 7803, N-5020 Bergen, Norway}
\emailAdd{pmmferreira@fc.ul.pt}
\emailAdd{bohdan.grzadkowski@fuw.edu.pl}
\emailAdd{omo@hvl.no}
\emailAdd{Per.Osland@uib.no}
\date{\today}

\abstract{Soft breaking of a symmetry requires an invariance of the dimension-4 part of the Lagrangian and non-trivial variation of the lower-dimensional part. However, in general, separation between the dim-4 and lower-dimensional Lagrangian is not invariant with respect to basis transformations of fields. Therefore, a natural question of the physical meaning of soft symmetry breaking arises. This problem is addressed here in the framework of two-Higgs-Doublet Models (2HDM). It has been shown, within these models, that in spite of the ambiguity corresponding to the separation between dim-4 and the lower-dimension Lagrangian, implications of the soft symmetry breaking could be formulated in terms of observables, i.e., they are physical and measurable. There are six global symmetries that can be imposed on the scalar sector of the generic 2HDM. Necessary and sufficient tree-level conditions for soft breaking of all of them have been formulated in terms of observables.}


\begin{document}

\maketitle

\flushbottom

\section{Introduction}
\label{sect:Introduction}
\setcounter{equation}{0}

The two-Higgs-doublet model (2HDM) is one of the simplest extensions of the Standard Model (SM) of particle
physics. It was introduced in 1973 by T.D.~Lee~\cite{Lee:1973iz} as a means to obtain an extra source, other than
the CKM matrix, of matter-antimatter asymmetry, via the spontaneous breaking of the  CP symmetry by vacuum expectation values (vevs) of Higgs doublets. To achieve this, the field content of the scalar sector of the model
is doubled relative to the SM -- instead of a single SU(2) doublet, the 2HDM has two. The model has a rich
phenomenology, with an enlarged scalar spectrum compared to that of the SM: two charged spin-0 particles and three
neutral ones. Certain versions of the model can provide dark matter candidates, or change the LHC-discovered
Higgs boson production and decay rates. For a review, see~\cite{Branco:2011iw}.

Early on it was recognised that the fact that both scalar doublets could couple independently to fermions
would induce scalar-mediated tree-level flavour-changing neutral currents (FCNC). Since these interactions
are severely constrained from numerous experimental observations in the quark and lepton sectors, they must
either be fine-tuned via judicious choices of Yukawa coupling values, suppressed by heavy-enough flavour-changing Higgs bosons, or set to zero naturally through the
imposition of a global symmetry on the model. The first such symmetry proposed was a $Z_2$ one, in a
 paper by Glashow and Weinberg \cite{Glashow:1976nt}, it forces a single scalar doublet to couple to
 fermions of the same electric charge, at the same time simplifying the 2HDM scalar potential.
 The global U(1) Peccei-Quinn symmetry~\cite{Peccei:1977hh} also eliminates tree-level FCNC, although
 leading to a different scalar potential. Said symmetries are imposed on the whole Lagrangian, but it is also of interest to find out what are the possible symmetries of the scalar potential alone. This question is made more complex by the fact that the two
doublets of the 2HDM carry the same quantum numbers, therefore any linear combination thereof which preserves their kinetic terms
is as physically acceptable as any other. This {\it basis freedom} can make it difficult to recognize
whether a given 2HDM scalar sector has a specific symmetry, since different field bases will imply
different symmetry-imposed relations among parameters of the potential.

Using a bilinear
formalism~\cite{Velhinho:1994np,Ivanov:2006yq,Maniatis:2006fs,Nishi:2006tg,Ivanov:2007de,Nishi:2007nh,
Maniatis:2007vn}, Ivanov was able to show \cite{Ivanov:2006yq,Ivanov:2007de} that
there are only
six different global symmetries one can impose upon the SU(2)$\times$U(1) 2HDM scalar potential.
Out of these six, three are so-called {\it Higgs family symmetries} -- unitary transformations
in which the doublets $\Phi_1$ and $\Phi_2$ mix among themselves, two of which are the aforementioned $Z_2$
and U(1) symmetries, the third one an SO(3) symmetry. The remaining three are {\it generalized CP
symmetries}, called CP1, CP2 and CP3, anti-unitary transformations which relate the doublets and
their complex conjugates. Ivanov proved this result by analysing the basis-invariant eigenvalues of
a matrix built from quartic couplings. Using a slightly different bilinear formalism, a more complete
set of basis-independent conditions for the presence of global symmetries in the 2HDM was explored in
Ref.~\cite{Ferreira:2010yh}. These were obtained in terms of properties of two vectors, together with the
eigenvalues and eigenvectors of a $3\times3$ matrix, constructed from the parameters of the potential.
This procedure, however, is not fully satisfactory, since the most general 2HDM potential contains redundant
parameters: of a total of 14 parameters, only 11 are necessary for a parametrization of  physics contained
in the scalar sector of the model~\cite{Davidson:2005cw}. Indeed, a set ${\cal P}$, of 11 physical parameters has been proposed \cite{Grzadkowski:2014ada,Grzadkowski:2016szj,Grzadkowski:2018ohf,Ogreid:2018bjq}, consisting of scalar masses,
couplings of scalars to gauge bosons and scalar trilinear and quartic interaction couplings --
to parameterize the 2HDM scalar potential. Conditions for CP conservation, for instance, were obtained
using this formalism.

Recently, we rephrased the symmetry conditions of~\cite{Ferreira:2010yh} in terms of the set ${\cal P}$ of
physical parameters \cite{Ferreira:2020ana}. Each symmetry -- in decreasing number of potential parameters,
CP1, $Z_2$, U(1), CP2, CP3 and SO(3) -- implies restrictions on the number of scalar
sector parameters, by imposing relationships among them, or eliminating some altogether. This, in terms of
the physical parameters in ${\cal P}$, translates as relationships among physical parameters, {\it i.e},
observables. However, unlike the relationships among potential parameters of~\cite{Ivanov:2007de}
or~\cite{Ferreira:2010yh}, the parameters in ${\cal P}$ are physical and correspond to quantities which
appear after spontaneous symmetry breaking, {\it e.g.,} scalar masses.
Such symmetry classification must take into account the vacuum state of the model, and whether that vacuum preserves the symmetry under consideration.

As a simple example, consider a model with a CP1-symmetric potential. For certain values of
its free parameters, the model may have a vacuum which preserves CP1, but in other regions of parameter
space the vacuum may spontaneously break that symmetry\footnote{However, no simultaneous CP1-preserving/
CP1-breaking minima may coexist in the potential for the same set of
parameters~\cite{Ferreira:2004yd,Barroso:2005sm}.}. In the former vacuum, two of the neutral scalars $H_i$
couple to gauge bosons via triple vertices of the form $H_i VV$, the third does not; in the latter vacuum,
all three neutral scalars have such couplings. The work of~\cite{Ferreira:2020ana} considered all possible
vacua present for each of the six exact symmetries, and listed the sets of relations between physical
observables of ${\cal P}$ for those vacua. Along the way it was discovered that many of the symmetry conditions
of~\cite{Ferreira:2010yh} amounted to tree-level fine tunings of parameters, which are not preserved under
renormalization -- for example, some conditions of \cite{Ferreira:2010yh} implied tree-level mass degeneracies
which were not preserved under radiative corrections, and as such those symmetry conditions amounted to
 tree-level accidents -- if a symmetry is present in the potential and imposes relations among parameters,
 those relations ought to be preserved under renormalization.

Our recent work \cite{Ferreira:2020ana} was however limited to symmetries of the potential as a whole (allowing for spontaneous breaking, but not for soft breaking). There is, on the
other hand, considerable interest in softly-broken 2HDM symmetries. For instance, models with an exact
$Z_2$ symmetry, spontaneously broken by the vacuum, predict an upper bound of about $\sim$ 800 GeV on masses
of the non-SM-like scalars, due to constraints from unitarity. Hence such models cannot have a {\it decoupling
limit} and have a considerably harder time fitting existing experimental data from the LHC. The introduction
of a dimension-2 term in the potential, which softly breaks the $Z_2$ symmetry (therefore preserving all
$Z_2$ conditions in the quartic couplings to all orders of perturbation theory) solves this issue, allowing
the extra scalars to have masses as high as necessary to accommodate all current LHC experimental results.

In this work we formulate and investigate tree-level constraints implied by softly broken symmetries of the 2HDM, i.e., when dim-4 interactions (quartic terms) are invariant while dim-2 (quadratic) are not.
Note that, in general, separation between dim-4 and the lower-dimensional Lagrangian is not invariant with respect to basis transformations of the fields.
For instance, in a case of softly broken $Z_2$ symmetry, $\lambda_6=\lambda_7=0$ and $m_{12}^2\neq 0$ in a certain basis. One can, however, perform a change of basis and rotate into a basis in which  $m_{12}^2 = 0$. In this new basis, the quartic couplings will change, and in particular $\lambda_{6,7}$ will become non-zero. It is then natural to question if the requirement of invariant quartic and modified quadratic terms will have the same physical consequences when formulated in those two different bases.
Therefore, a natural question of the physical significance of soft symmetry breaking arises. A straightforward strategy that might be used to clarify this issue is to adopt sufficient and necessary conditions for soft  breaking in a general basis. Since transformation rules of potential parameters for 2HDM are known, see e.g., \cite{Gunion:2005ja}, one could express the conditions in terms of parameters in another basis and this way explicitly verify if the conditions have the same form in terms of redefined parameters.  This is, however, not the procedure we have adopted.

Our strategy, altogether more ambitious, is to attempt to express the conditions for soft symmetry breaking in terms of invariant (with respect to basis transformations) observable quantities like tree-level masses and  appropriate tree-level couplings. If this procedure were to be successful, not only would a physical meaning of soft symmetry breaking be identified, but one could also formulate a strategy to test the breaking experimentally. This is the strategy we are opting for here.
Therefore the main goal of this work is to express the soft-symmetry breaking conditions in terms of physical parameters/observables like masses and couplings. Having masses and couplings measured, one can then relatively easily test the breaking of a given symmetry experimentally. As we will see in next sections, the constraints are somewhat complicated when expressed in terms of the physical parameters. However, the result is unique, i.e., there is no alternative (e.g., simpler) form of those constraints for a given set of observable input parameters.

In order to gain understanding and develop some intuition for the constraints, we will also formulate the results in the alignment limit, defined by requiring that couplings of the SM-like Higgs boson to be exactly as predicted by the SM. Applying this constraint, the symmetry conditions simplify, thus offering a chance for a slightly deeper understanding of the model.

This work is an extension of the study presented in \cite{Ferreira:2020ana}. There, we only studied the exact symmetries of the potential. In this work we allow for soft breaking of the six
symmetries. Besides purely theoretical curiosity explained above, our motivation for this study is the considerable interest for soft symmetry breaking in search for an optimal scalar sector of the electroweak theory. As before, our intent is to identify relationships among physical parameters within each model with a softly-broken symmetry.
We will only deal with the scalar and gauge sector of the theory -- the symmetries considered
could of course be extended to the fermion sector, but we will not deal with that issue here.
In fact, it is known that three of these symmetries (CP1, $Z_2$ and U(1)) have phenomenologically
valid fermionic extensions, one (CP3) can account for quark and charged lepton masses but seems numerically
unable to reproduce the CKM matrix~\cite{Ferreira:2010bm} and another (CP2) yields one generation of
massless charged fermions~\cite{Maniatis:2009vp,Maniatis:2009by,Ferreira:2010bm}. However, it does not seem to be impossible that all of them have hitherto
unknown valid fermionic extensions (for CP3, for instance, a 2HDM with a fermionic sector with
vector-like quarks is a phenomenologically valid model~\cite{Draper:2020tyq}).

This paper is organized as follows. In section~\ref{sect:The model} we define the model, together with our notation. Masses and couplings are presented, together with the employed physical parameter set. In section~\ref{sect:softsymm} we present conditions for the softly broken symmetries and relate them to potential parameters and the vacuum, in terms of physical parameters. In section~\ref{Sec:summary} we present a brief summary. Some technical material on quartic and quadratic invariants is collected in appendices \ref{App:I6Z} and \ref{App:bcal}. In appendix \ref{sec:soft-alignment} we explore how these softly broken symmetries look like in the Alignment Limit (AL). Finally, RG-unstable cases encountered are listed in appendix~\ref{App:nonRGE}.
\section{The model}
\label{sect:The model}
\setcounter{equation}{0}

We start out by parameterizing the scalar potential of the generic (CP-violating) 2HDM in the common fashion:
\begin{align}
V(\Phi_1,\Phi_2) &= -\frac12\left\{m_{11}^2\Phi_1^\dagger\Phi_1
+ m_{22}^2\Phi_2^\dagger\Phi_2 + \left[m_{12}^2 \Phi_1^\dagger \Phi_2
+ \hc\right]\right\} \nonumber \\
& + \frac{\lambda_1}{2}(\Phi_1^\dagger\Phi_1)^2
+ \frac{\lambda_2}{2}(\Phi_2^\dagger\Phi_2)^2
+ \lambda_3(\Phi_1^\dagger\Phi_1)(\Phi_2^\dagger\Phi_2) \nonumber \\
&+ \lambda_4(\Phi_1^\dagger\Phi_2)(\Phi_2^\dagger\Phi_1)
+ \frac12\left[\lambda_5(\Phi_1^\dagger\Phi_2)^2 + \hc\right]\nonumber\\
&+\left\{\left[\lambda_6(\Phi_1^\dagger\Phi_1)+\lambda_7
(\Phi_2^\dagger\Phi_2)\right](\Phi_1^\dagger\Phi_2)
+{\rm \hc}\right\}.
\label{Eq:pot}
\end{align}
The vacuum will be parameterized as
\bea
\langle\Phi_1\rangle=\frac{1}{\sqrt{2}}
\begin{pmatrix}
	0\\
	v_1
\end{pmatrix},\quad
\langle\Phi_2\rangle=\frac{e^{i\xi}}{\sqrt{2}}
\begin{pmatrix}
	0\\
	v_2
\end{pmatrix}.
\eea
All parameters in (\ref{Eq:pot}) are real, except for $m_{12}^2$, $\lambda_5$, $\lambda_6$ and $\lambda_7$,
which in general could be complex. Counting real and imaginary parts of each coupling, the potential of
Eq.~\eqref{Eq:pot} would seem to have 14 parameters, but in fact that number is reduced to 11 by the
{\it basis freedom} the potential has. In fact, since both doublets have identical quantum numbers, any linear
combination of them which preserves the $\Phi_{1,2}$ kinetic terms is as acceptable as any other, and this
allows some simplification of~\eqref{Eq:pot}.

 In order to see this explicitly, let us define $\bar\Phi_i=U_{ij}\Phi_j$, where $U$ is a generic U(2) matrix.
 If we express the potential in terms of $\{\bar\Phi_1\,,\,\bar\Phi_2\}$ instead of $\{\Phi_1\,,\,\Phi_2\}$,
 this is
 referred to as a change of basis. Note that the parameters of the potential will in general change under a change
 of basis. This is explicitly given in, for instance, Eqs. (5)--(15) of \cite{Gunion:2005ja}. Basis changes
 may therefore be used to express the potential in simpler terms, since sometimes it is possible to altogether
 eliminate some of its parameters. For instance, it is easy to find a U(2) matrix such that the
 terms of the 2HDM potential which are quadratic in the doublets are ``diagonalised" -- {\it i.e.,}
 we go to a basis where the new value of the $m^2_{12}$ coefficient is {\it zero}, thus eliminating two
 parameters out of the initial 14. In the new basis the quartic couplings $\lambda_5$, $\lambda_6$ and
 $\lambda_7$ will in general be complex, with unrelated phases -- but since in this basis $m^2_{12} = 0$, it is
 always possible to rephase one of the doublets so as to eliminate one of the complex phases of those
 three quartic couplings. For instance, if $\theta_5 = \arg(\lambda_5)$, the rephasing
 $\bar\Phi_2 = e^{i\theta_5/2} \Phi_2$ will render, in the new basis, $\lambda_5$ real, thus eliminating one
 more parameter from the potential, leaving us with 11 independent real parameters.

 \subsection{The physical parameter set}
\label{sec:paramset}

 The most general 2HDM thus needs 11 independent real parameters to fully characterize it.
Clearly, physics cannot depend on an arbitrary choice of basis for the Higgs doublets. All measurable quantities
must be basis independent, thereby leading to the study of basis invariant quantities.  Of course, the
scalar masses are basis invariant. The same holds for most of the physical couplings, though couplings involving charged fields may have arbitrary phases, and are thus
 pseudo-invariants~\cite{Grzadkowski:2014ada}. In recent works \cite{Grzadkowski:2014ada,Grzadkowski:2016szj,Grzadkowski:2018ohf} a set of 11 physical parameters was
 proposed to describe the 2HDM. That set includes, to begin with, the four masses of the scalar particles of
 the model -- three neutral states and a charged one. Those masses can obviously be
 related to the potential parameters in any basis, but for some applications it is simpler to work in the
 Higgs basis \cite{Donoghue:1978cj,Georgi:1978ri}. To reach a Higgs basis we perform a field rotation such that only
 one doublet has a non-vanishing, real and positive vacuum expectation value (VEV), whereas the other doublet has a vanishing VEV\footnote{The Higgs basis is not unique, as it is still possible to perform a basis change consisting of a U(1) rotation on $\Phi_2$, staying within the Higgs basis.}, i.e.
\begin{equation}
\langle \Phi_1 \rangle=\frac{1}{\sqrt{2}}\left(
\begin{array}{c}0\\
v
\end{array}\right),\quad
\langle \Phi_2 \rangle =\left(
\begin{array}{c}0\\
0
\end{array}\right),
\end{equation}
with $v=246\, {\rm GeV}$. We must make sure that the vacuum corresponds to a minimum of the potential, and by demanding that the derivatives of the potential with respect to the fields should vanish we end up with the stationary-point equations\footnote{Demanding that the vacuum should correspond to a stationary point does not guarantee that it is a minimum of the potential. One must also demand that the squared masses of the physical scalars are positive in order for the potential to have the curvature of a minimum point. In the present study we shall also encounter situations where some physical scalar has vanishing mass. Then we shall relax the requirement of positive squared masses to just demanding that the physical scalars have non-negative squared masses.}
\begin{equation}
m_{11}^2=v^2\lambda_1,\quad
m_{12}^2=v^2\lambda_6.
\end{equation}
We are assuming that a minimum of the potential can be found -- this is ensured by requiring that the potential
be {\em bounded from below} (BFB), so that no field direction exists for which the potential tends to minus infinity.
These BFB conditions \cite{Deshpande:1977rw} impose constraints on the quartic couplings, such as
demanding that $\lambda_1>0$ and $\lambda_2>0$. It has been shown, for the most general 2HDM potential,  that BFB
conditions are basis-invariant -- in the sense that, if a potential is BFB in one basis, it will be so in any
other~\cite{Ivanov:2006yq,Ivanov:2007de}. We are also assuming that this vacuum is the global minimum of the model, which in some cases
might not necessarily be true -- for instance, for the the softly broken $Z_2$ potential, there is the possibility of coexistence
of two minima which both spontaneously break $Z_2$~\cite{Ivanov:2007de,Barroso:2012mj,Barroso:2013awa}. This feature, however,
occurs only for very constrained regions of the parameter space. And even if the minima we choose in the current
work were local and not global, that would have no impact on the relations between physical parameters we will establish.
We should also mention that the globality of the vacuum is of course a basis-invariant statement, since it concerns the value of
the potential, which is basis-invariant by definition.

In the Higgs basis we may parameterize the two doublets as
\begin{equation}
\Phi_1=\left(
\begin{array}{c}G^+\\ (v+\eta_1+iG^0)/\sqrt{2}
\end{array}\right), \quad
\Phi_2=\left(
\begin{array}{c}H^+\\ (\eta_2+i\eta_3)/\sqrt{2}
\end{array}\right).
\end{equation}
The great advantage of the Higgs basis is the fact that the would-be Goldstone bosons are all contained in the
first doublet, and the charged field is automatically identified as the upper component of the
second doublet. Indeed,
we identify $G^0$ and $G^\pm$ as the massless would-be Goldstone fields, whereas $H^\pm$ are the massive charged scalars.
The neutral fields $\eta_i$ are not mass eigenstates, we relate them to the mass eigenstate fields $H_i$
by an orthogonal rotation matrix $R$ as
\begin{equation} \label{Eq:R-def}
\begin{pmatrix}
H_1 \\ H_2 \\ H_3
\end{pmatrix}
=R
\begin{pmatrix}
\eta_1 \\ \eta_2 \\ \eta_3
\end{pmatrix}.
\end{equation}
In general, the $H_{1,2,3}$ neutral mass eigenstates have no definite CP properties. This would
depend on the type of vacuum adopted -- if it preserves CP, two of these states will be CP even,
the third CP odd; but if there is no CP conservation in the model all three states will have undefined
CP properties.
As for the charged sector, the masses of the charged scalars $H^\pm$ can be read directly off from the corresponding bilinear terms in the potential, and are found, in the Higgs basis, to be given by
\begin{eqnarray}
M_{H^\pm}^2=-\frac{m_{22}^2}{2}+\frac{v^2}{2} \lambda_3.\label{chargedmass}
\end{eqnarray}
As for the neutral sector, the squared mass matrix is (in the Higgs basis) given by
\begin{equation}
{\cal M}^2
=v^2
\begin{pmatrix}
\lambda_1 & \Re \lambda_6 & -\Im \lambda_6 \\
\Re \lambda_6 & \frac{1}{2}(\lambda_3+\lambda_4+\Re \lambda_5-\frac{m_{22}^2}{v^2}) & -\frac{1}{2}\Im \lambda_5 \\
-\Im \lambda_6
& -\frac{1}{2}\Im \lambda_5 & \frac{1}{2}(\lambda_3+\lambda_4-\Re \lambda_5-\frac{m_{22}^2}{v^2})
\end{pmatrix}.\label{neutralmassmatrix}
\end{equation}
Then, by using (\ref{Eq:R-def}) we get the masses of the neutral scalars
from the diagonalization
of the mass-squared matrix, $\mcal^2$,
\begin{equation}
\label{Eq:cal-M}
{\rm diag}(M_1^2,M_2^2,M_3^2)=R{\cal M}^2R^{\rm T}.
\end{equation}

Thus the four masses -- the charged one from Eq.~\eqref{chargedmass} and the three neutral ones from
Eq.~\eqref{Eq:cal-M} -- are the first four of the physical parameter set ${\cal P}$. Next we consider
three parameters which govern the interactions of the neutral scalars with the electroweak gauge bosons.
To study such interactions, we must consider the kinetic term of the Lagrangian,
\begin{equation} \label{Eq:gauge-IS}
\lcal_k=(D_\mu \Phi_1)^\dagger(D^\mu \Phi_1) + (D_\mu \Phi_2)^\dagger(D^\mu \Phi_2),
\end{equation}
where $D^\mu=\partial^\mu+\frac{ig}{2}\sigma_i W_i^\mu+i\frac{g^\prime}{2}B^\mu$, and the
gauge mass eigenstates are related to the gauge fields by
\begin{subequations}
\begin{alignat}{2}
W_1^\mu&=\frac{1}{\sqrt{2}}(W^{+\mu}+W^{-\mu}), &\quad
W_2^\mu&=\frac{i}{\sqrt{2}}(W^{+\mu}-W^{-\mu}), \\
W_3^\mu&=\cos\thetaW Z^\mu+\sin\thetaW A^\mu, &\quad
B^\mu&=-\sin\thetaW Z^\mu+\cos\thetaW A^\mu.
\end{alignat}
\end{subequations}
The couplings between gauge fields and neutral scalars can now be read off from the kinetic term,
\begin{subequations}
\begin{align}
{\rm Coefficient}\left(\lcal_k,Z^\mu \left[H_j \overleftrightarrow{\partial_\mu}
H_i\right]\right)&=\frac{g}{2v\cos\thetaW}\epsilon_{ijk}e_k,\\
{\rm Coefficient}\left(\lcal_k,H_i Z^\mu Z^\nu\right)&=\frac{g^2}{4\cos^2\thetaW}e_i\,g_{\mu\nu},\\
{\rm Coefficient}\left(\lcal_k,H_i W^{+\mu} W^{-\nu}\right)&=\frac{g^2}{2}e_i\,g_{\mu\nu}.
\end{align}
\end{subequations}
Notice how all interactions between the $H_i$ and the electroweak gauge bosons involve the quantities
$e_i$ -- if $H_1$ is the SM-like Higgs boson, for instance, $e_1$ would be related to the Higgs-gauge
bosons coupling modifier $\kappa_V$ used by the LHC experimental collaborations (see for instance
\cite{Khachatryan:2016vau}) by $\kappa_V = e_1^2/v^2$. These quantities are given, in the Higgs basis, by
\begin{eqnarray}
e_i&\equiv& vR_{i1}.
\end{eqnarray}
In a general basis, with two VEVs for the doublets and a different rotation matrix $R$, these coefficients
would be given by $e_i=v_1R_{i1}+v_2R_{i2}$, and are found to be explicitly invariant under a change of basis
\cite{Ogreid:2018bjq}. The $e_i$  coefficients satisfy a ``sum rule"
\bea
e_1^2+e_2^2+e_3^3=v^2,
\eea
which is a consequence of the unitarity of the rotation matrix.

Finally, the physical set $\cal{P}$ is completed with three trilinear scalar couplings and a quartic one.
The three trilinear $H_iH^+H^-$ couplings and the quadrilinear $H^+H^+H^-H^-$ coupling are chosen,
denoted $q_i$ and $q$, respectively. In the Higgs basis they are
\begin{eqnarray}
q_{i}&\equiv&{\rm Coefficient}(V,H_iH^+H^-)\label{eq:qi}
\nonumber\\
&=&v(R_{i1}\lambda_3+R_{i2}\Re \lambda_7-R_{i3}\Im \lambda_7),\\
q&\equiv&{\rm Coefficient}(V,H^+H^+H^-H^-)\label{eq:q}
\nonumber\\
&=&\frac{1}{2}\lambda_2,
\end{eqnarray}
where the $R_{ij}$ are elements of the rotation matrix $R$ of (\ref{Eq:cal-M}).
In a general basis one can show explicitly that these couplings are all basis independent quantities \cite{Ogreid:2018bjq}. If $H_1$ is the SM-like Higgs boson, for instance, $q_1$ would contribute to its
diphoton width, since in the 2HDM that process involves a charged Higgs triangle diagram and thus
the $H_1 H^+H^-$ trilinear coupling.

To summarize, then, the physical set $\cal{P}$ is composed of the mass of the charged scalar; the masses of
the three neutral scalars; the coefficients $e_i$ of the gauge couplings of the neutral scalars; the trilinear
couplings $q_i$ between neutral and charged scalars, $H_iH^+H^-$; and the quartic charged coupling $q$,
corresponding to the vertex $H^+H^+H^-H^-$, so that\footnote{This set is not unique, see \cite{Ginzburg:2015cxa}. Note that the couplings $e_i$ and $q_i$ have dimension of mass.}
\begin{equation} \label{Eq:pcal}
{\cal P}\equiv\{M_{H^\pm}^2,M_1^2,M_2^2,M_3^2,e_1,e_2,e_3,q_1,q_2,q_3,q\}.
\end{equation}
These are all tree-level masses and couplings.
All physical observables of the scalar sector are expressible in terms of these 11 parameters \cite{Grzadkowski:2014ada,Grzadkowski:2016szj,Grzadkowski:2018ohf}.

\subsection{2HDM symmetries}
\label{sect:symm}

Considering the 2HDM scalar potential invariant under the gauge symmetry $\text{SU(2)}_L\times \text{U(1)}_Y$, there are
six possible global symmetries that one can impose on the model, via unitary (Higgs family symmetries) or
anti-unitary (generalized CP symmetries) transformations
among the two doublets\footnote{If one disregards the hypercharge group U(1) and Yukawa couplings, the list of possible 2HDM symmetries is larger, and may be found in~\cite{Battye:2011jj,Pilaftsis:2011ed}. We will not be
considering those extra symmetries.}. The symmetry transformations corresponding to each symmetry can themselves have different forms
in different bases and still yield the same physical models, but there is a ``canonical" form for each.
The first symmetry, which causes the least amount of constraints on the model, is \cp1, corresponding to
the ``standard" CP transformation:
\bit
\item \cp1:
\beq
\Phi_1 \to \Phi_1^\star\;,\;\Phi_2 \to \Phi_2^\star .
\label{eq:cp1}
\eeq
\eit
This yields a potential which, in the basis where the symmetry transformation is as above, has all
couplings real, and 9 independent real parameters. Next we have the $\z2$ symmetry, where one of
the doublets is transformed to change sign, thereby eliminating terms with an odd number of that doublet
field:
\bit
\item $\z2$:
\beq
\Phi_1\to\Phi_1\;,\; \Phi_2 \to -\Phi_2\,.
\label{eq:Z2}
\eeq
\eit
This yields a potential with 7 independent real parameters, all of them, without loss of generality,
real. The Peccei-Quinn \u1 symmetry, corresponding to the doublet transformation
\bit
\item \u1:
\beq
\Phi_1\to e^{-i\theta}\Phi_1\;,\; \Phi_2\to e^{i\theta}\Phi_2\,,
\label{eq:U1}
\eeq
\eit
where $\theta$ is an arbitrary angle, further reduces the number of parameters to 6, eliminating a
quartic $\lambda_5$ coupling. We then have the CP2 symmetry, where the fields transform as
\bit
\item CP2:
\beq
\Phi_1 \to \Phi_2^\star\;,\;\Phi_2 \to -\Phi_1^\star ,\label{CP2}
\eeq
\eit
forcing several quadratic and quartic parameters to be related to one another and others to be zero,
yielding a total of 5 independent real parameters. The phenomenology yielded by this symmetry was
thoroughly studied in a series of papers by Maniatis, Nachtmann and collaborators~\cite{Maniatis:2009vp,Maniatis:2009by}.
In the scalar
sector, CP2 is a special example of the CP3 symmetry (much like $Z_2$ is a special case of U(1)),
where  the fields transform as
\bit
\item CP3:
\beq
\Phi_i \to X_{ij} \Phi_j^\star\, ,\;\;\;\mbox{with}\;\;\; X=\begin{pmatrix} \cos\theta &
\sin\theta \\ -\sin\theta & \cos\theta \end{pmatrix}\,,\label{CP3}
\eeq
\eit
and $\theta$ is an arbitrary angle between $0$ and $\pi/2$ (the CP2 transformation corresponds to $\theta = \pi/2$, and the CP1 transformation corresponds to $\theta=0$).
This potential has only 4 independent real parameters and the CP3 symmetry has an interesting
extension to the fermionic sector~\cite{Ferreira:2010bm}. Finally, the most symmetry-constrained potential
is the one with an SO(3) symmetry, with field transformations given by
\bit
\item SO(3):
\beq
\Phi_i \to U_{ij} \Phi_j\;,\label{SO3}
\eeq
\eit
where $U$ is a generic unitary $2\times 2$ matrix. The SO(3) symmetry gives a potential with only
3 independent real parameters.

In table~\ref{tab:sym} we summarise the constraints on the parameters
of the potential of Eq.~\eqref{Eq:pot} of the ``canonical" form of these symmetry
transformations\footnote{Except for the CP2 symmetry, which we present in a simpler basis proposed
in~\cite{Davidson:2005cw}.} -- in other bases the symmetries would have different field transformations and correspond to different relationships between parameters, with no impact on the physics predicted by each model.

\begin{table}[h]
\begin{center}
\begin{tabular}
{|c|ccc|ccccccc|c|}
\hline \hline
 Symmetry &
$m_{11}^2$ & $m_{22}^2$ & $m_{12}^2$ &
$\lambda_1$ & $\lambda_2$ & $\lambda_3$ & $\lambda_4$ &
$\lambda_5$ & $\lambda_6$ & $\lambda_7$ & $N$ \\
\hline
 CP1 &  &  & real &
 & &  &  &
real & real & real & 9 \\
 $Z_2$ &   &   & 0 &   &  &  &  &
   & 0 & 0 & 7 \\
 U(1) &  &  & 0 &
 &  & &  &
0 & 0 & 0 & 6 \\
 CP2 &  & $m_{11}^2$ & 0 &
  & $\lambda_1$ &  &  &
  real & 0 &  0 & 5\\
 CP3 &  & $m_{11}^2$ & 0 &
   & $\lambda_1$ &  &  &
$\lambda_1 - \lambda_3 - \lambda_4$  & 0 & 0 & 4 \\
 SO(3) &  & $ m_{11}^2$ & 0 &
   & $\lambda_1$ &  & $\lambda_1 - \lambda_3$ &
0 & 0 & 0 & 3 \\
\hline \hline
\end{tabular}
\end{center}
\caption{Constraints on the 2HDM scalar potential parameters
due to each of the six symmetries, imposed using their ``canonical" forms (except for CP2).
In the final column we  show the total number $N$ of independent real parameters
for each symmetry-constrained scalar potential.
}
\label{tab:sym}
\end{table}

In ref.~\cite{Ferreira:2020ana} we identified the impact of each of these symmetries
in terms of relations among the physical parameters (\ref{Eq:pcal}). We now wish to study in detail
the possibility that these symmetries are not exact, but rather softly broken by dimension-2 terms.

\section{Softly broken symmetries}
\label{sect:softsymm}

A global symmetry is said to be {\it softly broken} when the
dim-4 part of the Lagrangian satisfies the symmetry constraints but parts of dim-2 or dim-3 ones do not.
The introduction of soft breaking terms in the model
-- terms quadratic or cubic in the fields, with coefficients of dimensions (mass)$^2$ or (mass)$^1$,
respectively -- does not spoil the renormalizability of the model~\cite{Symanzik:1969ek,Symanzik:1970zz}.
In other words, no new counterterms
need to be introduced in a model with a softly broken symmetry compared to the model with a potential where that symmetry
is intact. In practical terms, this means that we can have perfectly acceptable 2HDMs, from a renormalization
point of view, for which the quartic couplings obey the conditions laid out in table~\ref{tab:sym}, but the
quadratic ones do not. In fact, the most commonly studied version of the 2HDM is one with a softly broken $Z_2$
symmetry -- the symmetry would force the quadratic term $m_{12}^2$ to be zero, but this term is reintroduced
in the potential, so that a {\it decoupling limit}~\cite{Haber:1989xc,Gunion:2002zf} becomes accessible.
Indeed, the soft-breaking coefficient allows the masses of the extra scalars to be as high as desired, and
thus the model can easily comply with the so-far non-observation of new physics at the LHC.

Another basic
example is the Peccei-Quinn model, for which a massless axion appears if the symmetry is broken by the vacuum.
With a soft-breaking term $m_{12}^2$ in the potential, however, the axion becomes a massive pseudoscalar, with
mass proportional to $m_{12}^2$. The Minimal Supersymmetric Model, whose Higgs sector closely resembles the 2HDM,
also possesses a Peccei-Quinn symmetry which must be broken to avoid a massless axion, via the introduction of a soft-breaking term, the so-called $\mu$-term.

Notice a potential obstacle. Soft breaking of a symmetry requires an invariance of the dim-4 part of the Lagrangian and a non-trivial variation of the lower-dimensional part. However, in general, separation between the dim-4 and the lower-dimensional Lagrangian is not invariant with respect to basis transformations of fields. Therefore, a natural question of the physical meaning of soft symmetry breaking arises. It will be shown here that the conditions for soft symmetry breaking within the 2HDM can be formulated in terms of invariant (therefore, in principle, observable) quantities. Thus, in spite of the above-mentioned uncertainty, soft symmetry breaking does have a unique physical meaning.

It is therefore of great interest to study the possibility and implications of soft breakings of the six symmetries of the 2HDM.
As we will see, some symmetries may have different types of soft breaking, by leaving the potential invariant
to smaller symmetry groups. We will now go through the several possibilities for the 2HDM, and determine the
relations among the parameters of the  physical set $\cal{P}$ that they imply.

\subsection{Softly broken CP1 symmetry}
\label{sect:softCP1}

Demanding CP1 invariance in the quartic part of the potential implies $\Im \lambda_5=\Im \lambda_6=\Im
\lambda_7=0$, using the CP1 transformation of Eq.~\eqref{eq:cp1}. In the quadratic part this would render
$m^2_{12}$ real, as seen in table~\ref{tab:sym}. The only possibility of soft breaking, then, is to make this
parameter complex, which then  yields the potential for the most general 2HDM with a softly broken CP1 symmetry,
\begin{align}
	\label{Eq:softCP1pot}
	V(\Phi_1,\Phi_2) &= -\frac12\left\{m_{11}^2\Phi_1^\dagger\Phi_1
	+ m_{22}^2\Phi_2^\dagger\Phi_2 + \left[m_{12}^2 \Phi_1^\dagger \Phi_2
	+ \hc\right]\right\} \nonumber \\
	& + \frac{\lambda_1}{2}(\Phi_1^\dagger\Phi_1)^2
	+ \frac{\lambda_2}{2}(\Phi_2^\dagger\Phi_2)^2
	+ \lambda_3(\Phi_1^\dagger\Phi_1)(\Phi_2^\dagger\Phi_2)
	+ \lambda_4(\Phi_1^\dagger\Phi_2)(\Phi_2^\dagger\Phi_1)\nonumber \\
	&+ \frac{\Re\lambda_5}{2}\left[(\Phi_1^\dagger\Phi_2)^2 + (\Phi_2^\dagger\Phi_1)^2\right]\nonumber\\
	& +\left\{\left[\Re\lambda_6(\Phi_1^\dagger\Phi_1)+\Re\lambda_7
	(\Phi_2^\dagger\Phi_2)\right]\left[(\Phi_1^\dagger\Phi_2)+(\Phi_2^\dagger\Phi_1)
	\right]\right\}.
\end{align}

The only complex parameter in the potential is therefore $m^2_{12}$, and there is no basis redefinition
that can eliminate that complex phase. The potential therefore is {\it explicitly CP violating}.
However, this potential is not the same as the most general 2HDM -- the fact that a basis exists for which
the quartic part of the potential is real distinguishes this case from the most general 2HDM, for which
no such basis exists. In fact, this model possesses a {\it soft explicit CP violation}\footnote{A 2HDM
with this property was discussed in ref.~\cite{Ferreira:2011xc}.}. Also, notice that this model has a total of
10 independent real parameters, unlike the
11 of the most general 2HDM potential. In fact, we can imagine a basis rotation which absorbs the complex phase
of $m^2_{12}$, making it reappear in the quartic part; the new quartic coefficients $\lambda_{5,6,7}$ become
therefore complex, but their phases are closely correlated. A further basis rotation can eliminate the
remaining real part of $m^2_{12}$, changing $\lambda_i$ but not ruining the fact that
the phases of these couplings are correlated. The resulting potential has 2 quadratic and 4 quartic real
parameters and 3 quartic complex ones, out of which there is a single independent phase, a total of 10
real parameters.

We shall now discuss the conditions for the quartic part of the 2HDM potential to be invariant under \cp1 in
terms of the parameter set ${\cal P}$. To recap, there exists a basis in which the quartic part of
the potential is invariant under the CP1 transformation of Eq.~\eqref{eq:cp1}, which implies that
$\Im \lambda_5=\Im \lambda_6=\Im \lambda_7=0$. If at the same time no such basis
exists in which $\Im m_{12}^2=0$, we have softly broken \cp1. In \cite{Gunion:2005ja}, the
authors constructed four 2HDM basis-invariant quantities using the parameters of the potential. They
proved that if all four of these invariants were zero the  potential would be explicitly CP-conserving.
Out of these four invariants only one, their quantity $I_{6Z}$, is defined solely in terms of
quartic couplings, the remaining ones involving also the quadratic parameters. The authors
of~\cite{Gunion:2005ja} show, in their section~III, that the vanishing of $I_{6Z}$ is a necessary-and-sufficient condition
for the existence of a basis wherein the quartic part of the potential is real (therefore, CP1 invariant),
regardless of what happens with the quadratic terms. Using the techniques described in \cite{Ogreid:2018bjq},
$I_{6Z}$ can be expressed in terms of the parameter set ${\cal P}$. This conversion was
presented in an earlier work \cite{Grzadkowski:2016szj}. The result is recast here in Appendix~\ref{App:I6Z}.

Still, even with $I_{6Z} = 0$, the quadratic part of the potential may or may not be simultaneously \cp1
invariant, meaning \cp1 may or may not be softly broken. In order to guarantee a softly broken \cp1
symmetry, we must therefore complement the vanishing of $I_{6Z}$ with conditions preventing the quadratic part
of the potential from simultaneously becoming \cp1 invariant. While it is feasible to express such conditions in
terms of masses and couplings, it turns out to be impractical and not so transparent, therefore we choose
another approach. In an earlier work \cite{Ferreira:2020ana}, we presented conditions for the whole potential
to be \cp1 invariant, in terms of ${\cal P}$, and labelled them Case \textoverline{A}, \textoverline{B}, C and D\footnote{See Appendix~\ref{App:I6Z} for the definition of the quantities ${\Delta m_+}$ and ${\Delta q}$.},
\begin{alignat}{2}
	&\text{\bf Case \textoverline{A}:} &\ \
	&  M_1=M_2=M_3.  \nonumber \\
	&\text{\bf Case \textoverline{B}:} &\ \
	&  M_i=M_j,\quad (e_jq_i-e_iq_j)=0.  \nonumber \\
	&\text{\bf Case C:} &\ \
	&  e_k=q_k=0. \nonumber\\
	&\text{\bf Case D:} &\ \
	& {\Delta m_+}={\Delta q}=0.
	\nonumber
\end{alignat}
Those cases with a bar above were found to be RG-unstable \cite{Ferreira:2020ana} -- they in fact
correspond to a tree-level fine tuning, imposing relations among physical parameters (such as mass
degeneracies) which are not preserved under radiative corrections. We list those here for completeness,
but will, for the remainder of the paper, only focus on RG-stable symmetry conditions, excluding
possible fine tunings. Notice, however, that the two cases \textoverline{A} and \textoverline{B} both
involve mass degeneracy of neutral scalars. Thus, by requiring that no neutral scalars be degenerate in mass
we are effectively excluding the RG-unstable situations for CP1.
(Note, however, that mass degeneracies are allowed for some of the higher symmetries \cite{Ferreira:2020ana}.) Observe as well that Case C would imply the existence
of a neutral scalar $H_k$ which does not couple to electroweak gauge bosons nor to a pair of charged scalars -- {\it i.e.,} it is a pseudoscalar.

Thus, in order to guarantee a softly broken \cp1 (SOFT-CP1) we simply demand that there should not exist
relations among the physical parameters such that the conditions defining any of these four CP1-invariant
cases apply. Then, the criterion for having a softly broken \cp1 becomes
\begin{alignat}{2}
	&\text{\bf Case SOFT-CP1:} &\ \
	&  I_{6Z}=0 \text{ and none of the four cases of \cp1 invariant potential applies,}\nonumber \\
& & &	\text{for any combination of $i,j,k$.}\nonumber
\end{alignat}
Note that the condition for
invariance of the quartic part of the potential, $I_{6Z}=0$, is a quadratic equation that allows for solutions
with respect to ${\Delta m_+}$ in terms of the remaining parameters. That way the charged Higgs boson mass,
for instance, is no longer a free parameter and could be determined in terms of other parameters. Details
are presented in Appendix \ref{App:I6Z}. And since the physical set $\cal{P}$ has 11 parameters, the condition
$I_{6Z}=0$ reduces that number to 10, thus conforming to the parameter counting we had established earlier.
The expression for $I_{6Z}$ is quite unwieldy, being a homogeneous polynomial of order 6 in the lambdas\footnote{In terms of our parameter set $\pcal$, $I_{6Z}$ becomes a homogeneous polynomial of order 6 in the parameters $\{M_1^2, M_2^2, M_3^2, \mchsq, q_1, q_2, q_3, q\}$ with the gauge couplings $\{e_1, e_2, e_3\}$ acting as "coefficients".}, but it does
nonetheless force a relationship among the parameters of $\pcal$. From the expression for $I_{6Z}$ given in (\ref{I6Z}) we readily see that $I_{6Z}$ vanishes in Cases \textoverline{A}, \textoverline{B}, C and D (recalling that all $\Im J_i=0$, implying all $c_{ij}=0$ in Cases \textoverline{A}, \textoverline{B} and C).

\subsection{Softly broken \boldmath{$Z_2$} symmetry}
\label{sect:softZ2}

To softly break the $Z_2$ symmetry, the quartic parameters must obey the conditions laid out
in table~\ref{tab:sym} for that symmetry but the quadratic ones do not, which means that one
must reintroduce an $m_{12}^2$ coefficient in the potential.
Unlike the CP1 case, there is more than one way to softly break the $Z_2$ symmetry. Indeed there are three
different possibilities, yielding models with different phenomenologies. In fact, when the $Z_2$ symmetry is
softly broken, CP is no longer necessarily conserved, unlike the case when $Z_2$ is unbroken. And this CP
breaking can be either explicit or spontaneous.

The most general softly broken $Z_2$ potential, in the basis where the symmetry is defined as in Eq.~\eqref{eq:Z2}, is given by
\begin{align}
	\label{Eq:potZ2soft}
	V(\Phi_1,\Phi_2) &= -\frac12\left\{m_{11}^2\Phi_1^\dagger\Phi_1
	+ m_{22}^2\Phi_2^\dagger\Phi_2 +
	\left[
	m_{12}^2\Phi_1^\dagger\Phi_2+\text{H.c.}
	\right]\right\} \nonumber \\
	& + \frac{\lambda_1}{2}(\Phi_1^\dagger\Phi_1)^2
	+ \frac{\lambda_2}{2}(\Phi_2^\dagger\Phi_2)^2
	+ \lambda_3(\Phi_1^\dagger\Phi_1)(\Phi_2^\dagger\Phi_2)
	+ \lambda_4(\Phi_1^\dagger\Phi_2)(\Phi_2^\dagger\Phi_1)\nonumber \\
	&+ \frac12\left[\lambda_5(\Phi_1^\dagger\Phi_2)^2 + \hc\right],
\end{align}
where, other than $m_{12}^2$, all parameters are real. Three models are then possible:
\begin{itemize}
\item If the soft breaking parameter $m_{12}^2$ is complex then the potential has explicit
CP violation. This is the so-called {\it complex 2HDM}
(C2HDM)~\cite{Ginzburg:2002wt,Khater:2003wq,ElKaffas:2006gdt,ElKaffas:2007rq,WahabElKaffas:2007xd,Osland:2008aw,Grzadkowski:2009iz,Arhrib:2010ju},
with a total of 9
independent real parameters.
\item If the soft breaking parameter $m_{12}^2$ is real but the vacuum breaks CP, this model
has {\it spontaneous CP violation}, and the corresponding potential has 8 independent real
parameters.
\item If the soft breaking parameter $m_{12}^2$ is real but the vacuum preserves CP, we obtain
the most commonly studied version of the 2HDM, with the same number of parameters as the previous
case.
\end{itemize}

First, we present conditions for the quartic part of the 2HDM-potential to be invariant under $Z_2$ in terms of
our parameter set ${\cal P}$. If there exists a basis in which the quartic part of the potential is invariant
under the ``canonical" transformation of Eq.~\eqref{eq:Z2}, we say that the quartic part of the potential
is $Z_2$ invariant. This is equivalent to saying that there exists a basis in which $\lambda_6=\lambda_7=0$.
If at the same time no basis exists such that $m_{12}^2=0$, we have softly broken $Z_2$.
Basis-invariant conditions for $Z_2$ invariance were deduced in ref.~\cite{Davidson:2005cw}.
We are here interested in conditions concerning only the quartic sector. These involve the vanishing of a
commutator, which we call $\bcal$, between two $2\times 2$ matrices built from quartic couplings,
first introduced in Section III.B of \cite{Davidson:2005cw}. Again, using the techniques described in
\cite{Ogreid:2018bjq}, $\bcal$ can be expressed in terms of the parameter set ${\cal P}$.
Explicit calculations are shown in Appendix~\ref{App:bcal}, where we can see that $\bcal = 0$ introduces
two relations among the parameters of ${\cal P}$.

In our earlier work \cite{Ferreira:2020ana}, we presented conditions for the whole potential to be $Z_2$
invariant in terms of ${\cal P}$, listing a total of six cases. Four of the cases were found to be RG unstable, and we will not repeat them here. In fact, all such RG-unstable cases involved some mass degeneracy between neutral scalars. This
means that, considering always the general case where no neutral scalars are mass degenerate,
we are automatically avoiding the RG-unstable cases found in~\cite{Ferreira:2020ana} for $Z_2$. The two RG-stable
cases of exact $Z_2$ symmetry in the 2HDM potential are, using the notation of the work cited above,
\begin{alignat}{2}
	&\text{\bf Case CD:} &\ \
	&e_k=q_k=0, \quad
	2(e_j^2M_i^2+e_i^2M_j^2)M_{H^\pm}^2=v^2(e_jq_jM_i^2+e_iq_iM_j^2-M_i^2M_j^2),\nonumber\\
	& & & \phantom{\!\!\bullet}
	2(e_j^2M_i^2+e_i^2M_j^2)q=(e_jq_i-e_iq_j)^2+M_i^2M_j^2.\nonumber\\
	&\text{\bf Case CC:} &\ \
	&e_j=q_j=e_k=q_k=0.\nonumber
\end{alignat}
Case CD corresponds to a model where the $Z_2$ symmetry is spontaneously broken by the vacuum; case CC, on
the other hand, is a model for which the vacuum preserves $Z_2$, the so-called Inert Doublet Model (IDM) \cite{Ma:1978,Barbieri:2006,Cao:2007,LopezHonorez:2006}.
When accompanied by a soft breaking term $m_{12}^2\neq 0$,
the $Z_2$ symmetry could not be spontaneously broken anymore since it had already been explicitly broken. However still the vacuum could either be invariant or non-invariant under $Z_2$.

When working out the conditions for the vanishing of $\bcal$, in order to guarantee that $Z_2$ is in fact softly broken, we must in addition demand that there should not exist relations among
the physical parameters such that the constraints defining any of the six cases of a $\z2$-invariant potential apply. Here we list only two\footnote{In Appendix~\ref{App:bcal} we also found the RG-stable case SOFT-Z2-Y. It is not listed here since it is in fact only a special case of the more general case SOFT-CP2 which will be presented in Section \ref{sect:softCP2} on softly broken CP2.} RG-stable cases of softly broken $\z2$ (three more RG-unstable cases are listed in Appendix \ref{App:nonRGE} for completeness):

\begin{alignat}{2}
	&\text{\bf Case SOFT-Z2-X:} &\ \
	&\text{(\ref{Z2q}) and (\ref{Z2Mch}) apply, and none of the six cases of $\z2$-invariant}\nonumber \\
		& & &	\text{potential found in~\cite{Ferreira:2020ana} applies, for any combination of $i,j,k$. (There}\nonumber\\
			& & &\text{is an implicit assumption in (\ref{Z2q}) that $\Im J_1\neq0$.)}  \nonumber\\
	&\text{\bf Case SOFT-Z2-C:} &\ \
	&\text{$e_k=q_k=0$, $b_{ij}=0$, and none of the six cases of $\z2$-invariant }\nonumber \\
	& & &	\text{potential applies, for any combination of $i,j,k$.}  \nonumber
\end{alignat}
The quantity $b_{ij}$ is defined in (\ref{bij}).
The Case SOFT-Z2-C is CP conserving, and the last letter ``C" indicates that it is a sub-case of the \cp1-invariant case C presented in section~\ref{sect:softCP1}. We find that Case SOFT-Z2-X describes the C2HDM which is defined in terms of 9 parameters. Furthermore, the CP conserving case SOFT-Z2-C specifies a model where $m_{12}^2$ is real, the vacuum conserves CP, and the three constraints leave us with a model defined in terms of 8 free parameters.
\subsubsection{Softly broken \boldmath{$Z_2$} with spontaneous CP violation}
\label{sect:Z2spontaneous}
We have not yet identified all the constraints defining the model with a softly broken $\z2$, where $m_{12}^2$ is real and where CP is broken spontaneously. Such a model is contained within Case SOFT-Z2-X (the only CP-violating option), but these constraints have to be combined with the constraints of Case D of CP1, which will always happen whenever CP is violated spontaneously.

In order to have $\z2$-invariant quartic couplings, Eq.~(\ref{Z2q}) imposes a condition on $q$, whereas Eq.~(\ref{Z2Mch}) imposes a condition on $M_{H^\pm}^2$. In additions, there are conditions on $q$ and $M_{H^\pm}^2$ required for spontaneous CP violation, the ``Case~D conditions''. Superficially, it looks like this will leave us with four constraints. The two constraints on $q$ can be made compatible for $f_q=0$ and those on $M_{H^\pm}^2$ are compatible for $f_{m_+}=0$, where $f_q$ and $f_{m_+}$ are rational functions of the remaining 9 parameters of $\pcal$. Those conditions can be expressed as
\begin{equation}
f_q=g\times h_q=0, \quad f_{m_+}=g\times h_{m_+}=0.
\end{equation}
Thus, they are both satisfied for $g=0$, which is the condition (\ref{eq:spont}) below. The remaining factors, $h_q$ and $h_{m_+}$ can not simultaneously be zero, so there is no second solution. The new constraint we get ($g=0$) in addition to (\ref{Z2q}) and (\ref{Z2Mch}) is rather involved, it can be expressed in terms of the $d_{ijk}$ of Eq.~(\ref{Eq:d_ijk}) as
\begin{align}
&(2 d_{010} d_{012} d_{101}+d_{010} d_{012}^2+2 d_{010} d_{020}-3 d_{010} d_{022}-d_{010}^2 d_{101}+d_{010} d_{101}^2-d_{010} d_{200}\nonumber\\
&-2 d_{012} d_{020}-d_{012} d_{022}+d_{012} d_{200}+d_{020} d_{101}-2 d_{022} d_{101}-2 d_{030}+5 d_{032}-2 d_{101} d_{111}\nonumber\\
&+d_{210})\Im J_1
+(4 d_{010} d_{012}-2 d_{010}^2+2 d_{020}-4 d_{022})\Im J_{11}
\label{eq:spont}
\\
&+(d_{010} d_{012}-\frac{d_{010}^2}{2}+\frac{d_{020}}{2}-d_{022})\Im J_2
+(2 d_{010} d_{012}-d_{010}^2+d_{020}-2 d_{022})\Im J_{30}=0.\nonumber
\end{align}
This leaves us with
\begin{alignat}{2}
	&\text{\bf Case SOFT-Z2-XD:} &\ \
	&{\Delta m_+}={\Delta q}=0,\ \text{(\ref{eq:spont}) applies, and none of the six cases of $\z2$}\nonumber \\
	& & &	\text{invariant potential applies, for any combination of $i,j,k$.}  \nonumber
\end{alignat}
This case describes a model with soft breaking of $\z2$ and spontaneous CP violation. The number of parameters is again 8.

\subsection{Softly broken \boldmath{\u1} symmetry}
\label{sect:softU1}

The most general softly broken U(1) potential, in the basis where the symmetry is defined as in Eq.~\eqref{eq:U1}, is given by
\begin{align}
	\label{Eq:potU1soft}
	V(\Phi_1,\Phi_2) &= -\frac12\left\{m_{11}^2\Phi_1^\dagger\Phi_1
	+ m_{22}^2\Phi_2^\dagger\Phi_2 +
		m_{12}^2\left(\Phi_1^\dagger\Phi_2+\text{H.c.}
	\right)\right\} \nonumber \\
	& + \frac{\lambda_1}{2}(\Phi_1^\dagger\Phi_1)^2
	+ \frac{\lambda_2}{2}(\Phi_2^\dagger\Phi_2)^2
	+ \lambda_3(\Phi_1^\dagger\Phi_1)(\Phi_2^\dagger\Phi_2)
	+ \lambda_4(\Phi_1^\dagger\Phi_2)(\Phi_2^\dagger\Phi_1),
\end{align}
where all parameters are real. Notice that any complex phase of the soft breaking parameter $m_{12}^2$
may be absorbed by a trivial phase rotation of one of the fields, so we can consider, without
loss of generality, that it is real. As for the CP1 case, and unlike the $Z_2$ one, there is only
one possible form of soft breaking of the U(1) symmetry.
This is equivalent to saying that there exists a basis in which $\lambda_5=\lambda_6=\lambda_7=0$ but
no basis exists in which $m_{12}^2=0$ simultaneously.

From Table~II of \cite{Ferreira:2010yh} we see that the potential possesses a \u1 symmetry if one of the following two conditions is met:
\begin{alignat}{2}
	&\!\!\bullet &\ \
	&\Delta=0;\quad\text{and}\quad \vec{\xi}\times\vec{e}=\vec{\eta}\times\vec{e}=\vec{0}, \text{ where $\vec{e}$ is an eigenvector from a one-dimensional}\nonumber\\
	&\phantom{\!\!\bullet }&\ \
	&\text{eigenspace of } E, \\
	&\!\!\bullet &\ \
	&\Delta=0;\quad \Delta_0=0;\quad\text{and}\quad  \vec{\xi}\times\vec{\eta}=\vec{0}.
\end{alignat}
The constraint on $\vec{\xi}$ only affects the quadratic part of the potential, therefore the conditions for the quartic part of the potential to be \u1 invariant is
\begin{alignat}{2}
	&\!\!\bullet &\ \
	&\Delta=0;\quad\text{and}\quad \vec{\eta}\times\vec{e}=\vec{0}, \text{ where $\vec{e}$ is an eigenvector from a one-dimensional}\nonumber\\
	&\phantom{\!\!\bullet }&\ \
	&\text{eigenspace of } E, \text{ or}\\
	&\!\!\bullet &\ \
	&\Delta=0;\quad \Delta_0=0.
\end{alignat}
The vectors $\vec{\xi}$, $\vec{\eta}$, the matrix $E$ and the discriminants $\Delta$ and $\Delta_0$ were discussed and expressed in terms of masses and couplings in \cite{Ferreira:2020ana}, therefore we omit details here and only present the final results.
In \cite{Ferreira:2020ana}, we also presented conditions for the whole potential to be \u1 invariant in terms of ${\cal P}$, listing a total of four cases. Presenting only the two RG-stable cases, they correspond to
\begin{alignat}{2}
	&\text{\bf Case BCC:} &\ \
	&M_j=M_k,\quad
	e_j=q_j=e_k=q_k=0.\nonumber\\
	&\text{\bf Case C$_0$D:} &\ \
	&e_k=q_k=0, \quad 2(e_j^2M_i^2+e_i^2M_j^2)M_{H^\pm}^2=v^2(e_jq_jM_i^2+e_iq_iM_j^2-M_i^2M_j^2),\nonumber\\
	& & & \phantom{\!\!\bullet}
	2(e_j^2M_i^2+e_i^2M_j^2)q=(e_jq_i-e_iq_j)^2+M_i^2M_j^2,\quad M_k=0.\nonumber
\end{alignat}
Case BCC is the U(1) version of the IDM, where the global symmetry is preserved by the vacuum. Case
C$_0$D, on the other hand, corresponds to a vacuum for which the U(1) symmetry is spontaneously broken,
thus originating a massless scalar (an axion). With a soft breaking of
the symmetry, neither case is achievable -- given that both depend on the global U(1) being intact
before spontaneous symmetry breaking.

Thus, in order to guarantee a softly broken \u1, we simply demand that there should not exist
relations among the physical parameters such that the constraints defining any of the four cases of a U(1) invariant potential apply.
The RG-stable case is
\begin{alignat}{2}
	&\text{\bf Case SOFT-U1-C:} &\
	&e_k=q_k=0,\nonumber\\
	& & &2(e_j^2M_i^2+e_i^2M_j^2-v^2M_k^2)\mchsq\nonumber\\
	& & & =v^2((M_j^2-M_k^2)e_iq_i+(M_i^2-M_k^2)e_jq_j)\nonumber\\
	& & &\hspace*{0.5cm}+e_j^2(M_k^2-M_i^2)(M_j^2-2M_k^2)+e_i^2(M_k^2-M_j^2)(M_i^2-2M_k^2)
	,\nonumber\\
	& & & 2v^2(e_j^2M_i^2+e_i^2M_j^2-v^2M_k^2)q\nonumber\\
	& & &=v^2(e_iq_j-e_jq_i)^2+e_j^2M_j^2(M_i^2-M_k^2)+e_i^2M_i^2(M_j^2-M_k^2),\nonumber\\
	& & & \text{and none of the four cases of U(1) invariant potential }\nonumber \\
	& & & \text{found in \cite{Ferreira:2020ana} applies, for any combination of $i,j,k$.}\nonumber
\end{alignat}
This case is CP conserving, as it constitutes a subcase of Case C. An RG-unstable case is listed in Appendix \ref{App:nonRGE} for completeness.

\subsection{Softly broken \boldmath{CP2} symmetry}
\label{sect:softCP2}
The most general softly broken CP2 potential, in the basis where the symmetry is defined as in Eq.~\eqref{CP2}, is given by
\begin{align}
	\label{Eq:potCP2soft}
	V(\Phi_1,\Phi_2) &= -\frac12\left\{m_{11}^2\Phi_1^\dagger\Phi_1
	+ m_{22}^2\Phi_2^\dagger\Phi_2 + \left[m_{12}^2 \Phi_1^\dagger \Phi_2
	+ \hc\right]\right\} \nonumber \\
	&	+ \frac{\lambda_1}{2}\left\{(\Phi_1^\dagger\Phi_1)^2
	+ (\Phi_2^\dagger\Phi_2)^2\right\}
	+ \lambda_3(\Phi_1^\dagger\Phi_1)(\Phi_2^\dagger\Phi_2)
	+ \lambda_4(\Phi_1^\dagger\Phi_2)(\Phi_2^\dagger\Phi_1)\nonumber \\
	&+ \frac12\left[\lambda_5(\Phi_1^\dagger\Phi_2)^2 + \hc\right]
	+\left\{\lambda_6\left[(\Phi_1^\dagger\Phi_1)-
	(\Phi_2^\dagger\Phi_2)\right](\Phi_1^\dagger\Phi_2)
	+{\rm \hc}\right\}.
\end{align}
Without loss of generality we can employ a change of basis (making $\lambda_6=\lambda_7=0$ and $\lambda_5$ real) to get a simpler expression for the potential. Let's refer to this basis as the reduced CP2-basis in which the potential reads
\begin{align}
	\label{Eq:potCP2softreduced}
	V(\Phi_1,\Phi_2) &= -\frac12\left\{m_{11}^2\Phi_1^\dagger\Phi_1
	+ m_{22}^2\Phi_2^\dagger\Phi_2 + \left[m_{12}^2 \Phi_1^\dagger \Phi_2
	+ \hc\right]\right\} \nonumber \\
	&	+ \frac{\lambda_1}{2}\left\{(\Phi_1^\dagger\Phi_1)^2
	+ (\Phi_2^\dagger\Phi_2)^2\right\}
	+ \lambda_3(\Phi_1^\dagger\Phi_1)(\Phi_2^\dagger\Phi_2)
	+ \lambda_4(\Phi_1^\dagger\Phi_2)(\Phi_2^\dagger\Phi_1)\nonumber \\
	&+ \frac{\Re\lambda_5}{2}\left[(\Phi_1^\dagger\Phi_2)^2 + (\Phi_2^\dagger\Phi_1)^2\right]
	,
\end{align}
counting a total of 8 parameters.

If the quadratic part of the potential is not simultaneously invariant under the transformation (\ref{CP2}), we say that CP2 is softly broken by the potential. In the reduced CP2-basis this is equivalent to saying that we cannot simultaneously have $m_{12}^2=0$ and $m_{22}^2=m_{11}^2$.
From Table II of \cite{Ferreira:2010yh} we see that the potential possesses the CP2 symmetry \iff the following conditions are met:
\begin{alignat}{2}
	&\!\!\bullet &\ \
	&(\vec{\xi},\vec{\eta})=(\vec{0},\vec{0}).
\end{alignat}
The constraint on $\vec{\xi}$ only affects the quadratic part of the potential, thus the condition for the quartic part of the potential to be CP2 invariant becomes
\begin{alignat}{2}
	&\!\!\bullet &\ \
	&\vec{\eta}=\vec{0},
\end{alignat}
In \cite{Ferreira:2020ana}, we presented conditions for the whole potential to be CP2 invariant in terms of ${\cal P}$. Out of three cases, only one was found to be RG stable, namely
\begin{alignat}{2}
	&\text{\bf Case CCD:} &\ \
	&e_j=q_j=e_k=q_k=0,\quad 2M_{H^\pm}^2=e_iq_i-M_i^2,\quad 2v^2q=M_i^2.\nonumber
\end{alignat}
Thus, in order to guarantee a softly broken CP2, we simply demand that there should not exist relations among the physical parameters such that the constraints defining any of those three cases applies. Then, the conditions expressing a softly broken CP2 become
\begin{alignat}{2}
	&\text{\bf Case SOFT-CP2:} &\ \
	&v^2(e_1q_2-e_2q_1)+e_1e_2(M_2^2-M_1^2)=0,\nonumber\\
	& & &v^2(e_1q_3-e_3q_1)+e_1e_3(M_3^2-M_1^2)=0,\nonumber\\
	& & &v^2(e_2q_3-e_3q_2)+e_2e_3(M_3^2-M_2^2)=0,\nonumber\\
	& & & 2v^4q=e_1^2M_1^2+e_2^2M_2^2+e_3^2M_3^2,\nonumber\\
	& & & \text{and none of the three cases of CP2 invariant potential }\nonumber \\
	& & & \text{found in~\cite{Ferreira:2020ana} applies, for any combination of $i,j,k$.}\nonumber
\end{alignat}
These may look like four constraints, but the first three of these are not independent, so in reality these conditions only give
three constraints. This matches the number of free parameters, so there can be no more constraints in the general case. The authors of a recent paper \cite{Haber:2021zva} discuss the softly broken CP2 (naming it ERPS4). This model is fully described by the constraints in Case SOFT-CP2. The unbroken CP2 (their ERPS) is fully described by Case CCD.
\subsubsection{Distinguishing softly broken \boldmath{CP2} cases}
\label{sect:softCP2cases}
There are seemingly five different options for how to softly break CP2, namely
\begin{alignat}{4}
	&\!\!\text{Option I:} &\ \
	&m^2_{11} \neq m^2_{22} &,&\quad m^2_{12} \;\mbox{complex}, &\quad\text{(8 model parameters)}\nonumber\\
	&\!\!\text{Option II:} &\ \
	&m^2_{11} \neq m^2_{22} &,&\quad m^2_{12} \;\mbox{real}, &\quad\text{(7 model parameters)}\nonumber\\
	&\!\!\text{Option III:} &\ \
	&m^2_{11} \neq m^2_{22} &,&\quad m^2_{12}=0, &\quad\text{(6 model  parameters)}\nonumber\\
	&\!\!\text{Option IV:} &\ \
	&m^2_{11} = m^2_{22} &,&\quad m^2_{12} \;\mbox{complex}, &\quad\text{(7 model parameters)}\nonumber\\
	&\!\!\text{Option V:} &\ \
	&m^2_{11} = m^2_{22} &,&\quad m^2_{12} \;\mbox{real}, &\quad\text{(6 model parameters)}\nonumber
\end{alignat}
What distinguishes these different soft breaking options are the symmetries they leave unbroken. In fact, as in the $Z_2$ case,
different soft breakings will leave intact different ``pieces" of the original CP2 symmetry. The CP2-invariant potential in the reduced basis\footnote{Eq.~(\ref{Eq:potCP2softreduced}) with $m_{22}^2=m_{11}^2$ and $m_{12}^2=0$.}, can be conceived of
as resulting from the application of a $Z_2$ symmetry, followed by a permutation symmetry $S_2$, $\Phi_1\leftrightarrow \Phi_2$. Since the potential
in that basis is fully real, it is also trivially CP1 symmetric. The options for soft symmetry breaking above, then, correspond to: (I) no symmetry (other than gauge) left unbroken; (II) potential with residual CP1 symmetry; (III) potential with residual $Z_2$ (and CP1) symmetry;
(IV) potential with residual CP1 symmetry; (V) potential with residual $S_2$ (and CP1) symmetry. However, only three of these options are
relevant, as two of them can be related to others via basis transformations. Indeed, the residual CP1 symmetry of option IV only
becomes visible if one changes to a new basis given by the transformation
\begin{equation}
\begin{pmatrix}
	\bar{\Phi}_1\\
	\bar{\Phi}_2
\end{pmatrix}=
\frac{1}{\sqrt{2}}
\begin{pmatrix}
1 & 1\\
-i & i
\end{pmatrix}
\begin{pmatrix}
\Phi_1\\
\Phi_2
\end{pmatrix}
\end{equation}
which renders $\bar{m}_{12}^2$ real. After this change of basis, $\bar{m}_{11}\neq\bar{m}_{22}$. This means that the change of basis has led us back to Option II. Therefore Option II and Option IV ultimately lead to the same physics, making one of these two options redundant.
Likewise, starting from the potential described by Option V, we may perform a change of basis, using the transformation
\begin{equation}
\begin{pmatrix}
	\bar{\Phi}_1\\
	\bar{\Phi}_2
\end{pmatrix}=
\frac{1}{\sqrt{2}}
\begin{pmatrix}
	1 & 1\\
	-1 & 1
\end{pmatrix}
\begin{pmatrix}
	\Phi_1\\
	\Phi_2
\end{pmatrix}
\end{equation}
to make $\bar{m}_{12}^2$ vanish. This takes us to the potential described by Option III. Therefore Option III and Option V ultimately lead to the same physics, making one of these two options redundant.

All these options will satisfy the constraints of case SOFT-CP2, but only Option I, which is the most general way to softly break CP2 is fully described by case SOFT-CP2 without any additional constraints.

Options II and IV both have 7 model parameters, and they lead to two sub-cases of case SOFT-CP2 which both are described by four constraints. These are
\begin{alignat}{2}
	&\text{\bf Case SOFT-CP2-C:} &\ \
	&v^2(e_iq_j-e_jq_i)+e_ie_j(M_j^2-M_i^2)=0,\nonumber\\
	& & &e_k=q_k=0,\quad  q=\frac{e_i^2M_i^2+e_j^2M_j^2}{2v^4}.\nonumber\\
	&\text{\bf Case SOFT-CP2-D:} &\ \
	&v^2(e_1q_2-e_2q_1)+e_1e_2(M_2^2-M_1^2)=0,\nonumber\\
	& & &v^2(e_1q_3-e_3q_1)+e_1e_3(M_3^2-M_1^2)=0,\nonumber\\
	& & &v^2(e_2q_3-e_3q_2)+e_2e_3(M_3^2-M_2^2)=0,\nonumber\\
	& & &M_{H^\pm}^2=\frac{v^2(e_1q_1M_2^2M_3^2 + e_2q_2 M_3^2 M_1^2 + e_3 q_3 M_1^2 M_2^2-M_1^2 M_2^2 M_3^2)}{2(e_1^2M_2^2M_3^2+e_2^2M_3^2M_1^2+e_3^2 M_1^2 M_2^2)}
	,\nonumber\\
	& & & q=\frac{e_1^2M_1^2+e_2^2M_2^2+e_3^2M_3^2}{2v^4}.\nonumber
\end{alignat}
Case SOFT-CP2-C is what we get if Case SOFT-CP2 is combined with Case C of a CP1 invariant potential. This describes a model which conserves CP, and simultaneously softly breaks CP2. There are different ways to realize this case. For Option II, if we assume a vacuum where (in the ``reduced CP2 basis'') $\sin \xi=0$ and $v_2\neq v_1$, or for Option IV if we assume a vacuum where $v_2=v_1$, we will end up with a model fully described by Case SOFT-CP2-C.

Case SOFT-CP2-D is what we get if Case SOFT-CP2 is combined with Case D of a CP1 invariant potential. This describes a model which spontaneously breaks CP, and simultaneously softly breaks CP2. Note again that the number of constrains is in fact only four, since the first three of these constraints are not independent. There are again different ways to realize this case. For Option II, if we assume a vacuum where $\sin \xi\neq0$, or for Option IV if we assume a vacuum where $v_2\neq v_1$, we will end up with a model fully described by Case SOFT-CP2-D.

Options III and V both have 6 model parameters, and they lead to two sub-cases of case SOFT-CP2 which both are described by five constraints. These are
\begin{alignat}{2}
	&\text{\bf Case SOFT-CP2-CC:} &\ \
	&e_j=q_j=e_k=q_k=0,\quad  q=\frac{M_i^2}{2v^2}.\nonumber\\
	&\text{\bf Case SOFT-CP2-CD:} &\ \
	&v^2(e_iq_j-e_jq_i)+e_ie_j(M_j^2-M_i^2)=0,\nonumber\\
	& & &e_k=q_k=0,\quad q=\frac{e_i^2M_i^2+e_j^2M_j^2}{2v^4},\nonumber\\
	& & &M_{H^\pm}^2=\frac{v^2(e_iq_iM_j^2 + e_jq_j  M_i^2 -M_i^2 M_j^2 )}{2(e_i^2M_j^2+e_j^2M_i^2)}
.\nonumber
\end{alignat}
Case SOFT-CP2-CC is what we get if Case SOFT-CP2 is combined with Case CC of a $Z_2$ invariant potential. This describes a model which is $Z_2$ invariant, and simultaneously softly breaks CP2. In order to realize this case, we start from Option III and assume a vacuum where $v_2=0$ (or equivalently $v_1=0$), or we start from Option IV and assume both $v_2=v_1$ and $\sin\xi=0$ to end up with a model fully described by Case SOFT-CP2-CC.

Case SOFT-CP2-CD is what we get if Case SOFT-CP2 is combined with Case CD of a $Z_2$ invariant potential. This describes a model which spontaneously breaks $Z_2$, and simultaneously softly breaks CP2. There are again different ways to realize this case. For Option III, if we assume a vacuum where $\sin 2\xi=0$ and both $v_i\neq0$, or for Option V if we assume a vacuum where $\sin\xi=0$ and $v_2\neq v_1$ or a vacuum where $\sin\xi\neq0$ and $v_2= v_1$, we will end up with a model fully described by Case SOFT-CP2-CD.

\subsection{Softly broken \boldmath{CP3} symmetry}
\label{sect:softCP3}
The most general softly broken CP3 potential, in the basis where the symmetry is defined as in Eq.~\eqref{CP3}, is given by
\begin{align}
	\label{Eq:potCP3soft}
	V(\Phi_1,\Phi_2) &= -\frac12\left\{m_{11}^2\Phi_1^\dagger\Phi_1
	+ m_{22}^2\Phi_2^\dagger\Phi_2 + \left[m_{12}^2 \Phi_1^\dagger \Phi_2
	+ \hc\right]\right\} \nonumber \\
	&	+ \frac{\lambda_1}{2}\left\{(\Phi_1^\dagger\Phi_1)^2
	+ (\Phi_2^\dagger\Phi_2)^2\right\}
	+ \lambda_3(\Phi_1^\dagger\Phi_1)(\Phi_2^\dagger\Phi_2)
	+ \lambda_4(\Phi_1^\dagger\Phi_2)(\Phi_2^\dagger\Phi_1)\nonumber \\
	&+ \frac{\lambda_1-\lambda_3-\lambda_4}{2}\left[(\Phi_1^\dagger\Phi_2)^2 + (\Phi_2^\dagger\Phi_1)^2\right].
\end{align}
Without loss of generality we can employ a change of basis (making $m_{22}^2=m_{11}^2$) to get a simpler expression for the potential. Let's refer to this basis as the reduced CP3-basis in which the potential reads
\begin{align}
	\label{Eq:potCP3softreduced}
	V(\Phi_1,\Phi_2) &= -\frac12\left\{m_{11}^2\left(\Phi_1^\dagger\Phi_1
	+ \Phi_2^\dagger\Phi_2\right) + \left[m_{12}^2 \Phi_1^\dagger \Phi_2
	+ \hc\right]\right\} \nonumber \\
	&	+ \frac{\lambda_1}{2}\left\{(\Phi_1^\dagger\Phi_1)^2
	+ (\Phi_2^\dagger\Phi_2)^2\right\}
	+ \lambda_3(\Phi_1^\dagger\Phi_1)(\Phi_2^\dagger\Phi_2)
	+ \lambda_4(\Phi_1^\dagger\Phi_2)(\Phi_2^\dagger\Phi_1)\nonumber \\
	&+ \frac{\lambda_1-\lambda_3-\lambda_4}{2}\left[(\Phi_1^\dagger\Phi_2)^2 + (\Phi_2^\dagger\Phi_1)^2\right]
	,
\end{align}
counting a total of 6 parameters.
If the quadratic part of the potential is not simultaneously invariant under the same transformation, we say that CP3 is softly broken by the potential. In the reduced CP3-basis this is equivalent to saying that we cannot simultaneously have $m_{12}^2=0$.
From Table II of \cite{Ferreira:2010yh} we see that the potential possesses a CP3 symmetry \iff the following conditions are met:
\begin{alignat}{2}
	&\!\!\bullet &\ \
	&\Delta=0;\quad\text{with}\quad(\vec{\xi},\vec{\eta})=(\vec{0},\vec{0}).
\end{alignat}
The constraint on $\vec{\xi}$ only affects the quadratic part of the potential, so the condition for the quartic part of the potential to be CP3 invariant becomes
\begin{alignat}{2}
	&\!\!\bullet &\ \
	&\Delta=0;\quad\text{with}\quad\vec{\eta}=\vec{0}.
\end{alignat}
In \cite{Ferreira:2020ana}, we presented conditions for the whole potential to be CP3 invariant in terms of ${\cal P}$,
identifying a total of five cases of which two were found to be RG stable:
\begin{alignat}{2}
	&\text{\bf Case BCCD:} &\ \
	&e_j=q_j=e_k=q_k=0,\quad 2M_{H^\pm}^2=e_iq_i-M_i^2,\quad 2v^2q=M_i^2,\quad M_j=M_k.\nonumber\\
	&\text{\bf Case C$_0$CD:} &\ \
	&e_j=q_j=e_k=q_k=0,\quad 2M_{H^\pm}^2=e_iq_i-M_i^2,\quad 2v^2q=M_i^2,\quad M_j=0.\nonumber
\end{alignat}
Thus, in order to guarantee a softly broken CP3, we demand that there should not exist relations among the physical parameters such that the constraints defining any of those cases of a CP3 invariant potential applies. Then, the conditions expressing a softly broken CP3 become (listing only the RG-stable cases):
\begin{alignat}{2}
	&\text{\bf Case SOFT-CP3-B:} &\
	&M_j=M_k,\quad e_jq_k-e_kq_j=0,\nonumber\\
	& & &	v^2(e_iq_j-e_jq_i)+e_ie_j(M_j^2-M_i^2)=0,
	\nonumber\\
	& & & v^2(e_iq_k-e_kq_i)+e_ie_k(M_j^2-M_i^2)=0,\nonumber\\
	& & &	2v^4q=e_i^2M_i^2+(e_j^2+e_k^2)M_j^2,\quad \nonumber\\
	& & &2v^2\mchsq=e_i^2M_i^2+(e_j^2+e_k^2)M_j^2+v^2(e_1q_1+e_2q_2+e_3q_3),\nonumber\\
	& & & \text{and none of the five cases of CP3 invariant potential }\nonumber \\
	& & & \text{found in~\cite{Ferreira:2020ana} applies, for any combination of $i,j,k$.}\nonumber\\
	&\text{\bf Case SOFT-CP3-C:} &\
	&e_k=q_k=0,\quad v^2(e_iq_j-e_jq_i)+e_ie_j(M_j^2-M_i^2)=0,\quad\nonumber\\
	& & &
	2v^4q=e_i^2M_i^2+e_j^2M_j^2,\quad\nonumber\\
	& & & (e_j^2M_i^2+e_i^2M_j^2-v^2M_k^2)\nonumber\\
	& & &\times\left[2v^2\mchsq+e_i^2M_i^2+e_j^2M_j^2-v^2(2M_k^2+e_iq_i+e_jq_j)\right]\nonumber\\
	& & &=2e_i^2e_j^2(M_j^2-M_i^2)^2,\nonumber\\
	& & & \text{and none of the five cases of CP3 invariant potential }\nonumber \\
	& & & \text{found in~\cite{Ferreira:2020ana} applies, for any combination of $i,j,k$.}\nonumber
\end{alignat}
Case SOFT-CP3-B contains partial mass degeneracy, which requires $m_{11}^2+m_{22}^2=0$, whereas case SOFT-CP3-C represents the general case of softly broken CP3.
Another, RG-unstable case is listed in Appendix \ref{App:nonRGE} for completeness.

Then there is seemingly three different options for how to softly break CP3, namely
\begin{alignat}{4}
	&\!\!\text{Option I:} &\ \
	&m^2_{12} \;\mbox{complex}, &\quad\text{(6 model parameters)}\nonumber\\
	&\!\!\text{Option II:} &\ \
	&m^2_{12} \;\mbox{imaginary}, &\quad\text{(5 model parameters)}\nonumber\\
	&\!\!\text{Option III:} &\ \
	&m^2_{12} \;\mbox{real},  &\quad\text{(5 model  parameters)}\nonumber
\end{alignat}
All these options will satisfy the constraints of case SOFT-CP3-C, but only Option I, which is the most general way to softly break CP3 is fully described by case SOFT-CP3-C without any additional constraints.
\begin{alignat}{2}
	&\text{\bf Case SOFT-CP3-CC:} &\
	&e_j=q_j=e_k=q_k=0,\quad
	2v^2q=M_i^2,\quad\nonumber\\
	& & & \mchsq=\frac{1}{2}(2M_k^2-M_i^2+e_iq_i),\nonumber\\
	&\text{\bf Case SOFT-CP3-CD:} &\
	&e_k=q_k=0,\quad v^2(e_iq_j-e_jq_i)+e_ie_j(M_j^2-M_i^2)=0,\quad\nonumber\\
	& & &
	2v^4q=e_i^2M_i^2+e_j^2M_j^2,\quad\nonumber\\
	& & & 2(e_j^2M_i^2+e_i^2M_j^2)M_{H^\pm}^2=v^2(e_jq_jM_i^2+e_iq_iM_j^2-M_i^2M_j^2),\nonumber\\
	& & & e_i^2M_j^2(M_j^2-M_k^2)+e_j^2M_i^2(M_i^2-M_k^2)=0,\nonumber\\
	&\text{\bf Case SOFT-CP3-BCC:} &\
	&e_j=q_j=e_k=q_k=0,\quad M_j=M_k,\quad
	2v^2q=M_i^2,\nonumber\\
	&\text{\bf Case SOFT-CP3-C$_0$D:} &\
	&e_k=q_k=0,\quad M_k=0,\quad v^2(e_iq_j-e_jq_i)+e_ie_j(M_j^2-M_i^2)=0,\quad\nonumber\\
	& & &
	2v^4q=e_i^2M_i^2+e_j^2M_j^2,\quad\nonumber\\
	& & & 2(e_j^2M_i^2+e_i^2M_j^2)M_{H^\pm}^2=v^2(e_jq_jM_i^2+e_iq_iM_j^2-M_i^2M_j^2),\nonumber	
\end{alignat}
Case SOFT-CP3-CC is what we get if Case SOFT-CP3-C is combined with Case CC of a $Z_2$ invariant potential. This describes a model which is $Z_2$ invariant, and simultaneously softly breaks CP3. In order to realize this case, we start from Option III and assume a vacuum where $\sin\xi=0$ and $v_1=v_2$ to end up with a model fully described by Case SOFT-CP3-CC.

Case SOFT-CP3-CD is what we get if Case SOFT-CP3-C is combined with Case CD of a $Z_2$ invariant potential. This describes a model which spontaneously breaks $Z_2$, and simultaneously softly breaks CP3. In order to realize this case, we start from Option III and assume a vacuum where $\sin \xi\neq 0$ and $v_1=v_2$ to end up with a model fully described by Case SOFT-CP3-CD.
Note that the last of the constraints for case SOFT-CP3-CD comes from equating the expressions for $\mchsq$ from case SOFT-CP3-C and case D and working out the resulting condition.

Case SOFT-CP3-BCC is what we get if Case SOFT-CP3-C is combined with Case BCC of a $U(1)$ invariant potential. This describes a model which is $U(1)$ invariant, and simultaneously softly breaks CP3. In order to realize this case, we start from Option II and assume a vacuum where $\cos\xi\neq0$ or $v_1\neq v_2$ to end up with a model fully described by Case SOFT-CP3-BCC.

Case SOFT-CP3-C$_0$D is what we get if Case SOFT-CP3-C is combined with Case C$_0$D of a $U(1)$ invariant potential. This describes a model which spontaneously breaks $U(1)$, and simultaneously softly breaks CP3. In order to realize this case, we start from Option II and assume a vacuum where $\cos \xi= 0$ and $v_1=v_2$ to end up with a model fully described by Case SOFT-CP3-C$_0$D.

\subsection{Softly broken \boldmath{SO(3)} symmetry}
\label{sect:softSO3}
The most general softly broken SO(3) potential, in the basis where the symmetry is defined as in Eq.~\eqref{SO3}, is given by
\begin{align}
	\label{Eq:potSO3soft}
	V(\Phi_1,\Phi_2) &= -\frac12\left\{m_{11}^2\Phi_1^\dagger\Phi_1
	+ m_{22}^2\Phi_2^\dagger\Phi_2 + \left[m_{12}^2 \Phi_1^\dagger \Phi_2
	+ \hc\right]\right\} \nonumber \\
	&	+ \frac{\lambda_1}{2}\left\{(\Phi_1^\dagger\Phi_1)^2
	+ (\Phi_2^\dagger\Phi_2)^2\right\}
	+ \lambda_3(\Phi_1^\dagger\Phi_1)(\Phi_2^\dagger\Phi_2)
	+ (\lambda_1-\lambda_3)(\Phi_1^\dagger\Phi_2)(\Phi_2^\dagger\Phi_1).
\end{align}
Without loss of generality we can employ a change of basis (making $m_{12}^2=0$) to get a simpler expression for the quadratic part of the potential. The SO(3)-invariant quartic part of the potential remains unchanged under all basis changes. Let's refer to this simpler basis as the reduced SO(3)-basis in which the potential reads
\begin{align}
	\label{Eq:potSO3softreduced}
	V(\Phi_1,\Phi_2) &= -\frac12\left\{m_{11}^2\Phi_1^\dagger\Phi_1
	+ m_{22}^2\Phi_2^\dagger\Phi_2 \right\} \nonumber \\
	&	+ \frac{\lambda_1}{2}\left\{(\Phi_1^\dagger\Phi_1)^2
	+ (\Phi_2^\dagger\Phi_2)^2\right\}
	+ \lambda_3(\Phi_1^\dagger\Phi_1)(\Phi_2^\dagger\Phi_2)
	+ (\lambda_1-\lambda_3)(\Phi_1^\dagger\Phi_2)(\Phi_2^\dagger\Phi_1)
	,
\end{align}
counting a total of 4 parameters.
If the quadratic part of the potential is not invariant under the same transformation, we say that SO(3) is softly broken by the potential. In the reduced SO(3)-basis this is equivalent to saying that we cannot simultaneously have $m_{11}^2=m_{22}^2$.
From Table II of \cite{Ferreira:2010yh} we see that the potential possesses the SO(3) symmetry \iff the following conditions are met:
\begin{alignat}{2}
	&\!\!\bullet &\ \
	&\Delta=0;\quad \Delta_0=0;\quad\text{with}\quad(\vec{\xi},\vec{\eta})=(\vec{0},\vec{0}).
\end{alignat}
The constraint on $\vec{\xi}$ only affects the quadratic part of the potential, the condition for the quartic part of the potential to be SO(3) invariant is
\begin{alignat}{2}
	&\!\!\bullet &\ \
	&\Delta=0;\quad \Delta_0=0;\quad\text{with}\quad\vec{\eta}=\vec{0},
\end{alignat}
In \cite{Ferreira:2020ana}, we presented conditions for the whole potential to be SO(3) invariant, identifying two cases, of which one is RG stable, namely
\begin{alignat}{2}
	&\text{\bf Case B$_0$C$_0$C$_0$D:} &\
	&M_j=M_k=0,\quad e_j=q_j=e_k=q_k=0,\nonumber\\
	& & &2M_{H^\pm}^2=e_iq_i-M_i^2,\quad 2v^2q=M_i^2.\nonumber
\end{alignat}
Thus, in order to guarantee a softly broken SO(3), we demand that there should not exist relations among the physical parameters such that the constraints defining any of those two cases apply. Then, the conditions expressing a softly broken SO(3) become
\begin{alignat}{2}
	&\text{\bf Case SOFT-SO3-ABBB:} &\ \
	&M_1=M_2=M_3,\quad
	{\qcal}^2=0,\nonumber\\
	& & &2M_{H^\pm}^2=M_1^2+e_1q_1+e_2q_2+e_3q_3,\quad2v^2 q=M_1^2,\nonumber\\
	& & & \text{and none of the two cases of SO(3) invariant potential }\nonumber \\
	& & & \text{found in~\cite{Ferreira:2020ana} applies, for any combination of $i,j,k$.}\nonumber\\
	&\text{\bf Case SOFT-SO3-BCC:} &\
	&M_j=M_k,\quad e_j=q_j=e_k=q_k=0,\nonumber\\
	& & &2M_{H^\pm}^2=e_iq_i-M_i^2+2M_j^2,\quad 2v^2q=M_i^2,\nonumber\\
	& & & \text{and none of the two cases of SO(3) invariant potential }\nonumber \\
	& & & \text{found in~\cite{Ferreira:2020ana} applies, for any combination of $i,j,k$.}\nonumber
\end{alignat}
Both these two cases are found to be RG stable. Case SOFT-SO3-ABBB contains full mass degeneracy and requires $m_{11}^2+m_{22}^2=0$, whereas Case SOFT-SO3-BCC represents the general case of softly broken SO(3) and requires $v_2=0$ (or equivalently $v_1=0$) in the reduced SO(3)-basis.
\section{Summary}
\label{Sec:summary}
The weak-basis invariant formulation of soft symmetry breaking of the scalar 2HDM potential has been developed.
Soft breaking of a symmetry requires an invariance of the dimension-4 part of the Lagrangian and a non-trivial variation of the lower-dimensional part. In general, the dim-4 and the lower-dimensional parts of the potential change under a basis transformations of the Higgs doublets. It has been shown that in spite of the ambiguity corresponding to the separation between the dim-4 and the lower-dimensional Lagrangian, implications of the soft symmetry breaking could be formulated in terms of tree-level observables, i.e., they are physical and testable. There exist six global symmetries that one can impose on the scalar potential of the model. Necessary and sufficient conditions for soft breaking of all of them have been formulated for the generic 2HDM in terms of observables.

\begin{figure}[htb]
	\centering
	\includegraphics[scale=0.85]{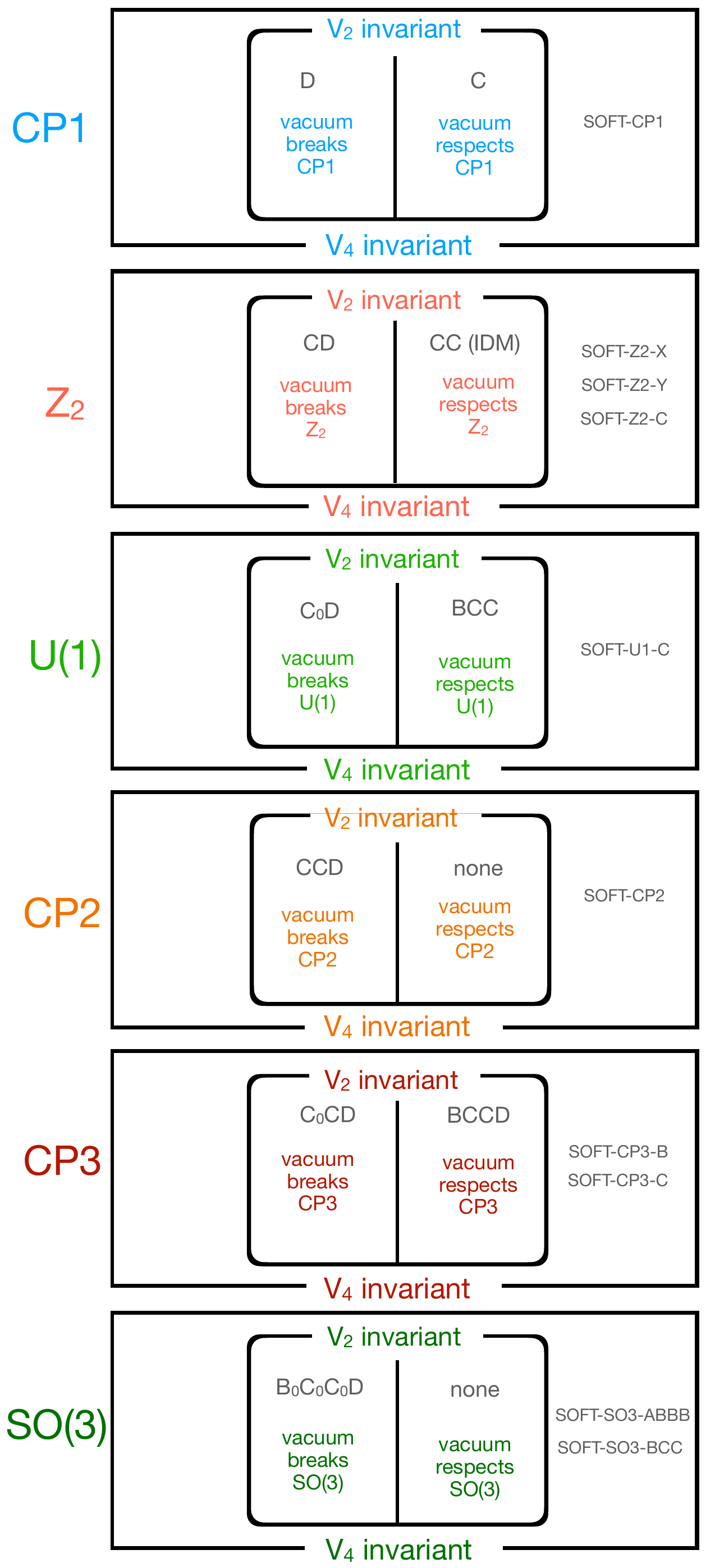}
	\caption{Overview of the six symmetries, with a listing of the different ways these can be broken. $V_{2,4}$ denote dim-2 and dim-4 parts of the scalar potential, respectively.}
	\label{Fig:tableau}
\end{figure}

In figure~\ref{Fig:tableau} we give an overview of the different ways the six symmetries can be broken. The least symmetric case is when the CP1 symmetry is softly broken. This case is referred to as ``SOFT-CP1'', and was discussed in section~\ref{sect:softCP1}. The different higher symmetries can also be softly broken, some in more than one way. Accordingly, in addition to the models with a fully symmetric potential \cite{Ferreira:2020ana}, there is a plethora of addtional models, in this paper denoted by a name whose prefix is ``SOFT''.

In order to gain some further understanding of the conditions for soft symmetry breaking we have also found their form in the alignment limit (AL), defined by requiring that couplings of the SM-like Higgs boson discovered at the LHC are exactly as predicted by the SM. The results for the AL are shown in Appendix~\ref{sec:soft-alignment}.

In table~\ref{Table:numbers} we list the number of free parameters that are compatible with the different symmetries.  Note that two models with different symmetries may have the same number of free parameters.

We recall that all the relations between couplings and physical parameters shown in this paper were based on tree-level results obtained from the Lagrangian.
One could wonder if our symmetry-based  relations remain valid if radiative corrections were taken into account. To answer this question one would need to calculate the 1-loop (at least) effective potential, define and determine physical masses and couplings and then find potential parameters ($m_{ij}^2$ and $\lambda_i$) in terms of physical (observable) parameters like masses and couplings. Those relations would receive, in general, loop corrections. However, even though this would be an attractive project, nevertheless it lies far beyond the scope of our work. Nevertheless a few remarks concerning the relevance of loop corrections are useful.

Let us consider, as an illustration, the model with a $Z_2$ symmetry imposed, which
forces the relation $\lambda_6 = \lambda_7 = 0$.
With or without soft breakings, the model is renormalizable, which means that couterterms $\delta \lambda_{6,7}$ are not needed to get rid of all divergences.  As a consequence of that the relation $\lambda_6 = \lambda_7 = 0$ is preserved by RG evolution to all orders of perturbation
expansion. However, if the $Z_2$ symmetry is softly broken, finite corrections to $\lambda_6$ and $\lambda_7$ are expected. Indeed, the
authors of~\cite{BhupalDev:2014bir} argue that such finite threshold effects might induce non-zero value for $\lambda_6$ and $\lambda_7$.

First let us emphasize the role played by a symmetry of vacuum, it turns out that it is as relevant as a soft breaking of the symmetry. We have found many examples in which  the vacuum leaves  intact certain symmetries after spontaneous symmetry breaking~\cite{Ferreira:2020ana}. For instance, if the whole potential is $Z_2$-invariant there are two possible models implying different physics. If only one doublet has a non-vanishing vev we get the inert doublet model, and for this model the $Z_2$ symmetry is exact. If both doublets have non-zero vevs, $Z_2$ is spontaneously broken. Both cases stem from a potential with the same symmetry, but the physics they imply is different.

For the inert doublet model, since the $Z_2$ symmetry is preserved after spontaneous symmetry breaking, the relation $\lambda_6 = \lambda_7 = 0$ will hold at all orders, also when considering finite contributions from radiative corrections to these couplings. But if the vacuum spontaneously
breaks $Z_2$, though no infinite contributions to $\lambda_6$ and $\lambda_7$ will ever appear in a loop calculation, finite contributions
to those couplings may well emerge. It is important to stress that in this case one can expect $\lambda_6$ and $\lambda_7$ to be much smaller than remaining (tree-level allowed) quartic couplings $\lambda_{1-5}$.
If $\lambda_{6,7}$ were experimentally confirmed in the future to be indeed  much smaller than other quartic couplings it would indicate the case of spontaneously broken $Z_2$ rather than a model with tree-level allowed $\lambda_6$ and $\lambda_7$.
One could then conclude that the non-zero values of $\lambda_6$ and $\lambda_7$ were not independent of the rest of the parameters, but rather follow the expected loop relations deduced from a model with spontaneouly broken $Z_2$, retaining the number of free parameters of the model.

The presence of soft breaking terms would manifest itself at higher orders in much the same manner as discussed above. If one were to consider a potential
with $Z_2$ symmetry softly broken by, for instance, a real $m^2_{12}$, one may expect to find finite contributions to $\lambda_6$ and $\lambda_7$
appearing in loop calculations, now depending on $m^2_{12}$, as presumed in eq. (3.17) of ref.~\cite{BhupalDev:2014bir}. But the number of free
parameters of the model remains the same (in this case 8), and the non-zero values of $\lambda_6$ and $\lambda_7$ are not independent, but rather follow
the formulae deduced from a loop calculation in the softly broken $Z_2$ model. And that would suggest that the model had an underlying, softly broken, symmetry.\footnote{Or, spontaneously broken $Z_2$.} Furthermore, they can be expected to be much smaller than the
other quartic couplings, so that the tree-level expressions used throughout this paper should constitute a useful approximation to study of the model. The same argument also holds for all the other symmetries discussed in this work.
Concluding, it might be interesting to study relations between couplings and physical parameters beyond the tree level approximation. However, since radiative corrections do not imply symmetry breaking~\footnote{In this work we limit ourself to tree-level approximation. There are, however, cases where loop contributions are indispensable. Anomalous symmetries, like e.g. the scale invariance may serve as an illustration. In this case the invariance is broken by 1-loop corrections and masses are generated, see \cite{Coleman:1973jx}. However, we do not consider anomalous symmetries here.} we do not expect them to change the tree-level predictions drastically, while definitely it would fall outside the scope of the current work.

\clearpage

\begin{table}[htb]
\begin{center}
\begin{tabular}{|c||c|c|c|c|c|c|c|}
\hline
\hline
& CP1 & $Z_2$ & U(1) & CP2 & CP3 & SO(3)\\
\hline\hline
unbroken  & 9 & 7 & 6 & 5 & 4 & 3\\
\hline
soft breaking & 10 & 9 & 7 & 8 & 6  & 4 \\
\hline
\end{tabular}
\end{center}
\caption{\label{Table:numbers} Free parameters (not counting inequalities) in models with various symmetries. Only the numbers for the most general model containing each symmetry is given.}
\end{table}

\vspace*{10mm}
\noindent
{\bf Acknowledgments.}
P.M.F.'s research is supported
by \textit{Funda\c c\~ao para a Ci\^encia e a Tecnologia} (FCT)
through contracts
UIDB/00618/2020, UIDP/00618/2020, CERN/FIS-PAR/0004/2019, CERN/FIS-PAR/0014/2019.
The work of B.G. is supported in part by the National Science Centre (Poland) as a research project, decision no 2017/25/B/ST2/00191 and 2020/37/B/ST2/02746.
The research of P.O. has been supported in part by the Research Council of Norway.
\appendix

\section{The invariant $I_{6Z}$ in terms of couplings and masses}
\label{App:I6Z}

We will need the following useful dimensionless abbreviations \cite{Grzadkowski:2016szj},
\begin{align} \label{Eq:d_ijk}
d_{ijk}&=\frac{q_{1}^i M_1^{2j}e_1^k+q_{2}^i M_2^{2j}e_2^k+q_{3}^i M_3^{2j}e_3^k}{v^{i+2j+k}},\\
m_+&=\frac{M_{H^\pm}^2}{v^2}, \quad
m_i=\frac{M_i^2}{v^2}, \label{Eq:def-m1}
\end{align}
and also the $CP$-odd invariants
\begin{eqnarray}
	\Im J_1&=&\frac{1}{v^5}\sum_{i,j,k}\epsilon_{ijk}M_i^2e_ie_kq_j\nonumber\\
	 &=&\frac{1}{v^5}[e_1e_2q_3(M_2^2-M_1^2)-e_1e_3q_2(M_3^2-M_1^2)+e_2e_3q_1(M_3^2-M_2^2)],\label{eq:imjJ-physical}\\
	\Im J_2&=&\frac{2}{v^9}\sum_{i,j,k}\epsilon_{ijk}e_ie_je_kM_i^4M_k^2=\frac{2e_1 e_2 e_3}{v^9}\sum_{i,j,k}\epsilon_{ijk}M_i^4M_k^2\nonumber\\
	&=&\frac{2e_1 e_2 e_3}{v^9}(M_2^2-M_1^2)(M_3^2-M_2^2)(M_3^2-M_1^2),\label{eq:imJ2-physical}\\
	\Im J_{30}&\equiv&\frac{1}{v^5}\sum_{i,j,k}\epsilon_{ijk} q_i M_i^2  e_j q_k,\nonumber\\
	 &=&\frac{1}{v^5}[q_1q_2e_3(M_2^2-M_1^2)-q_1q_3e_2(M_3^2-M_1^2)+q_2q_3e_1(M_3^2-M_2^2)]\label{eq:imJ3-physical},\nonumber\\
	\Im J_{11}&\equiv&\frac{1}{v^7}\sum_{i,j,k}\epsilon_{ijk}e_i M_i^2 M_j^2 e_k q_j \\
	 &=&\frac{1}{v^7}[e_1e_2q_3M_3^2(M_2^2-M_1^2)-e_1e_3q_2M_2^2(M_3^2-M_1^2)+e_2e_3q_1M_1^2(M_3^2-M_2^2)].\nonumber\\
\end{eqnarray}

The expression for $I_{6Z}$, whose vanishing guarantees a \cp1 invariant quartic potential is large and unwieldy when expressed directly in terms of the masses and couplings of ${\cal P}$. Therefore we introduce some abbreviations,
let us first introduce the notation
\bea
{\Delta m_+}&\equiv&\frac{\mchsq}{v^2}-\tilde{m}_+,\\
{\Delta q}&\equiv&q-\tilde{q},
\eea
where
\bea
\tilde{m}_+&\equiv&\frac{e_1q_1M_2^2M_3^2+e_2q_2M_1^2M_3^2+e_3q_3M_1^2M_2^2-M_1^2M_2^2M_3^2}{2(e_1^2M_2^2M_3^2+e_2^2M_3^2M_1^2+e_3^2 M_1^2 M_2^2)},\\
\tilde{q}&\equiv&\frac{(e_2q_3-e_3q_2)^2M_1^2+(e_3q_1-e_1q_3)^2M_2^2+(e_1q_2-e_2q_1)^2M_3^2+M_1^2M_2^2M_3^2}{2(e_1^2M_2^2M_3^2+e_2^2M_3^2M_1^2+e_3^2 M_1^2 M_2^2)}.
\eea
Using this notation, we find	
\begin{eqnarray}
	I_{6Z}&=&c_{21}{(\Delta m_+)}^2{\Delta q}+c_{20}{(\Delta m_+)}^2+c_{12}{\Delta m_+}{(\Delta q)}^2+c_{11}{\Delta m_+}{\Delta q}+c_{10}{\Delta m_+}\nonumber\\
	&&+c_{03}{(\Delta q)}^3+c_{02}{(\Delta q)}^2+c_{01}{\Delta q},\label{I6Z}
\end{eqnarray}
where $c_{ij}$ are polynomials in terms of model parameters, given by
\bea
c_{21}&=& 16\left(- d_{010}+ d_{012}+ d_{101}\right)\Im J_1+32 \Im J_{11}+8 \Im J_2+16 \Im J_{30},\\
c_{20}&=&\left(2 \tilde{q}-d_{012}\right) \left( 8\left(- d_{010}+ d_{012}+ d_{101}\right)\Im J_1+16 \Im J_{11}+4 \Im J_2+8 \Im J_{30}\right),\\
c_{12}&=&16 \left(d_{012}-d_{010}\right)\Im J_1+32 \Im J_{11}+16 \Im J_2,\\
c_{11}&=&
8  \left(-4 d_{010} d_{012}-4 \tilde{m}_+ d_{010}-4 \tilde{q} d_{010}+2 d_{010}^2+2 d_{012} d_{101}\right.\nonumber\\
&&\left. \hspace*{0.5cm}+4 \tilde{m}_+ d_{012}+4 \tilde{q} d_{012}+d_{012}^2+d_{022}+4 \tilde{m}_+ d_{101}-d_{101}^2-d_{200}\right)\Im J_1\nonumber\\
&&+32  \left(-d_{010}+d_{012}-d_{101}+2 \tilde{m}_++2 \tilde{q}\right)\Im J_{11}\nonumber\\
&&+8  \left(-d_{010}-d_{101}+2 \tilde{m}_++4 \tilde{q}\right)\Im J_2\nonumber\\
&&+16  \left(-d_{010}+2 d_{012}-d_{101}+2 \tilde{m}_+\right)\Im J_{30}
,\\
c_{10}&=&-4 \left(-4 \tilde{m}_+ d_{010} d_{012}+8 \tilde{q} d_{010} d_{012}+2 d_{010}^2 d_{012}-5 d_{010} d_{022}-2 d_{010} d_{101}^2\right.\nonumber\\
&&\left.\hspace*{0.8cm}+2 d_{010} d_{200}+8 \tilde{m}_+ \tilde{q} d_{010}+4 \tilde{q}^2 d_{010}-4 \tilde{q} d_{010}^2+4 \tilde{m}_+ d_{012} d_{101}\right.\nonumber\\
&&\left.\hspace*{0.8cm}-4 \tilde{q} d_{012} d_{101}+d_{012} d_{101}^2-3 d_{012} d_{200}-8 \tilde{m}_+ \tilde{q} d_{012}+4 \tilde{m}_+ d_{012}^2\right.\nonumber\\
&&\left.\hspace*{0.8cm}-4 \tilde{q}^2 d_{012}-2 \tilde{q} d_{012}^2+2 d_{022} d_{101}-2 \tilde{q} d_{022}+3 d_{032}+2 d_{101} d_{111}\right.\nonumber\\
&&\left.\hspace*{0.8cm}-2 d_{101} d_{200}-8 \tilde{m}_+ \tilde{q} d_{101}+2 \tilde{q} d_{101}^2+2 d_{101}^3+2 \tilde{q} d_{200}-2 d_{210}\right) \Im J_1\nonumber\\
&&+4 \left(4 d_{010} d_{012}-8 \tilde{q} d_{010}+4 d_{012} d_{101}-8 \tilde{m}_+ d_{012}+8 \tilde{q} d_{012}+d_{012}^2\right.\nonumber\\
&&\left.\hspace*{0.8cm}-7 d_{022}-8 \tilde{q} d_{101}-2 d_{101}^2+2 d_{200}+16 \tilde{m}_+ \tilde{q}+8 \tilde{q}^2\right)\Im J_{11}\nonumber\\
&&+2 \left(2 d_{010} d_{012}-4 \tilde{q} d_{010}-d_{012} d_{101}-4 \tilde{m}_+ d_{012}-2 d_{022}-4 \tilde{q} d_{101}\right.\nonumber\\
&&\left.\hspace*{0.8cm}-2 d_{101}^2+3 d_{111}+2 d_{200}+8 \tilde{m}_+ \tilde{q}+8 \tilde{q}^2\right)\Im J_2\nonumber\\
&&+8 \left(d_{010} d_{012}-2 \tilde{q} d_{010}-2 \tilde{m}_+ d_{012}+4 \tilde{q} d_{012}-d_{012}^2\right.\nonumber\\
&&\left.\hspace*{0.8cm}-d_{022}-2 \tilde{q} d_{101}-d_{101}^2+d_{111}+d_{200}+4 \tilde{m}_+ \tilde{q}\right)\Im J_{30},\\
c_{03}&=&8\Im J_2,\\
c_{02}&=&-8  \left(5 d_{010} d_{012}-d_{010} d_{101}+2 \tilde{m}_+ d_{010}\right.\nonumber\\
&&\left. \hspace*{0.8cm}-2 d_{010}^2-2 \tilde{m}_+ d_{012}-5 d_{022}+d_{111}+2 d_{020}\right)\Im J_1\nonumber\\
&&+16  \left(-d_{010}+2 d_{012}-d_{101}+2 \tilde{m}_+\right)\Im J_{11}\nonumber\\
&&+4 \left(-2 d_{010}+d_{012}-2 d_{101}+4 \tilde{m}_++6 \tilde{q}\right) \Im J_2,\\
c_{01}&=&-4  \left(5 d_{010} d_{012} d_{101}+8 \tilde{m}_+ d_{010} d_{012}+20 \tilde{q} d_{010} d_{012}+d_{010} d_{012}^2\right.\nonumber\\
&&\left. \hspace*{0.8cm}+2 d_{010}^2 d_{012}+4 d_{010} d_{020}-10 d_{010} d_{022}-4 \tilde{q} d_{010} d_{101}\right.\nonumber\\
&&\left. \hspace*{0.8cm}-3 d_{010} d_{101}^2-3 d_{010} d_{111}+2 d_{010} d_{200}+4 \tilde{m}_+^2 d_{010}+8 \tilde{m}_+ \tilde{q} d_{010}\right.\nonumber\\
&&\left. \hspace*{0.8cm}-4 \tilde{m}_+ d_{010}^2-8 \tilde{q} d_{010}^2-8 d_{012} d_{020}-d_{012} d_{022}-4 \tilde{m}_+ d_{012} d_{101}\right.\nonumber\\
&&\left. \hspace*{0.8cm}+d_{012} d_{111}-d_{012} d_{200}-4 \tilde{m}_+^2 d_{012}-8 \tilde{m}_+ \tilde{q} d_{012}-2 \tilde{m}_+ d_{012}^2\right.\nonumber\\
&&\left. \hspace*{0.8cm}+2 d_{020} d_{101}+8 \tilde{q} d_{020}-6 d_{022} d_{101}-2 \tilde{m}_+ d_{022}-20 \tilde{q} d_{022}-4 d_{030}\right.\nonumber\\
&&\left. \hspace*{0.8cm}+16 d_{032}+4 d_{101} d_{111}-d_{101} d_{200}\right.\nonumber\\
&&\left. \hspace*{0.8cm}-4 \tilde{m}_+^2 d_{101}+2 \tilde{m}_+ d_{101}^2+4 \tilde{q} d_{111}+2 \tilde{m}_+ d_{200}\right)\Im J_1\nonumber\\
&&+8  \left(-4 d_{010} d_{012}+2 d_{010} d_{101}-4 \tilde{m}_+ d_{010}-4 \tilde{q} d_{010}+2 d_{010}^2\right.\nonumber\\
&&\left. \hspace*{0.8cm}+d_{012} d_{101}+4 \tilde{m}_+ d_{012}+8 \tilde{q} d_{012}-2 d_{020}+3 d_{022}-4 \tilde{m}_+ d_{101}\right.\nonumber\\
&&\left. \hspace*{0.8cm}-4 \tilde{q} d_{101}-2 d_{101}^2-3 d_{111}+3 d_{200}+4 \tilde{m}_+^2+8 \tilde{m}_+ \tilde{q}\right)\Im J_{11}\nonumber\\
&&+2  \left(-2 d_{010} d_{012}+2 d_{010} d_{101}-4 \tilde{m}_+ d_{010}-8 \tilde{q} d_{010}\right.\nonumber\\
&&\left. \hspace*{0.8cm}+2 d_{010}^2+4 \tilde{q} d_{012}-2 d_{020}+2 d_{022}-4 \tilde{m}_+ d_{101}\right.\nonumber\\
&&\left. \hspace*{0.8cm}-8 \tilde{q} d_{101}-4 d_{101}^2+5 d_{200}+4 \tilde{m}_+^2+16 \tilde{m}_+ \tilde{q}+12 \tilde{q}^2\right)\Im J_2\nonumber\\
&&+4  \left(-6 d_{010} d_{012}+2 d_{010} d_{101}-4 \tilde{m}_+ d_{010}+2 d_{010}^2+8 \tilde{m}_+ d_{012}\right.\nonumber\\
&&\left. \hspace*{0.8cm}-2 d_{020}+7 d_{022}-4 \tilde{m}_+ d_{101}+d_{101}^2-4 d_{111}+4 \tilde{m}_+^2\right)\Im J_{30}.
\eea
There are different ways to realize $I_{6Z}=0$.
\begin{enumerate}
	\item We easily see that $I_{6Z}$ vanishes when all $\Im J_i$ vanish. This corresponds to a CP1-invariant potential (both the quadratic and quartic parts) as well as a CP-invariant vacuum. There is no CP-violation in the scalar sector.
	\item We easily see that $I_{6Z}$ also vanishes whenever ${\Delta m_+}={\Delta q}=0$. We know that this corresponds to a CP1-invariant potential (both the quadratic and quartic parts). If at least one $\Im J_i$ is nonzero, we have spontaneous CP  violation.
	\item Situations under which $I_{6Z}$ vanish, other than the two mentioned above, yield models with softly broken CP1.
\end{enumerate}

\section{The commutator $\bcal$}
\label{App:bcal}

It is convenient to rewrite the scalar potential of Eq.~\eqref{Eq:pot} as
\begin{equation}
V\equiv
Y_{a\bar{b}}\Phi_{\bar{a}}^\dagger\Phi_b+\frac{1}{2}Z_{a\bar{b}c\bar{d}}(\Phi_{\bar{a}}^\dagger\Phi_b)(\Phi_{\bar{c}}^\dagger\Phi_d).
\label{Eq:pot-other}
\end{equation}
In this second form, a summation over barred with un-barred indices is implied, e.g., $a=\bar a=1,2$. Thus,
\bea \label{Eq:cap_Y}
Y_{11}=-\frac{m_{11}^2}{2},\quad
Y_{12}=-\frac{m_{12}^2}{2},\quad  Y_{21}=-\frac{(m_{12}^2)^*}{2},\quad
Y_{22}=-\frac{m_{22}^2}{2}
\eea
and
\bea
&&Z_{1111}=\lambda_1,\quad Z_{2222}=\lambda_2,\quad Z_{1122}=Z_{2211}=\lambda_3,\nonumber\\
&&Z_{1221}=Z_{2112}=\lambda_4,\quad Z_{1212}=\lambda_5,\quad Z_{2121}=(\lambda_5)^*,\nonumber\\
&&Z_{1112}=Z_{1211}=\lambda_6,\quad Z_{1121}=Z_{2111}=(\lambda_6)^*,\nonumber\\
&&Z_{1222}=Z_{2212}=\lambda_7,\quad Z_{2122}=Z_{2221}=(\lambda_7)^*.\label{Zlambdas}
\eea
All other $Z_{a\bar{b}c\bar{d}}$ vanish.

In Section III.B of \cite{Davidson:2005cw}, three commutators of $2\times2$ matrices are presented, whose simultaneous vanishing guarantees a $Z_2$ invariant potential. For our purposes, we shall need only one of these commutators in order to guarantee a $Z_2$ invariant quartic potential, namely
\bea
\bcal &\equiv& [Z^{(1)},Z^{(11)}],
\eea
where
\bea
Z_{a\bar{d}}^{(1)}&=&Z_{a\bar{b}b\bar{d}},\\
Z_{c\bar{d}}^{(11)}&=&Z_{ba}^{(1)}Z_{a\bar{b}c\bar{d}},
\eea
are given in terms of the potential coefficients in (\ref{Zlambdas}).
Let us point out that $\bcal$ is not basis-invariant, nor are its elements. However, the condition $\bcal=0$ (requiring all four matrix elements to vanish simultaneously) is a basis-invariant constraint.
We translate the elements of $\bcal$ into masses and couplings by working in the Higgs basis using the method described in \cite{Ogreid:2018bjq}, yielding the following matrix-elements of $\bcal$,
\bea
\bcal_{11}&=&2i\left[(2q-2d_{010}+d_{012}+2d_{101})\Im J_1 +4\,\Im J_{11}+\Im J_2+2\,\Im J_{30}\right],\label{B11}\\
\bcal_{12}&=&\frac{1}{v^6}\sum_{i=1}^3\left[
2  \left(d_{012}-2 q\right)M_i^4e_i
+2 v^2 \left(d_{012}-2 q\right)M_i^2q_i\right.\nonumber\\
&&\hspace*{1.3cm}+v^2 \left[d_{012} \left(2 q-d_{101}\right)+m_+ \left(4 q-2 d_{012}\right)\right.\nonumber\\
&&\hspace*{2.1cm}\left.+2 \left(-d_{022}+q d_{101}-d_{101}^2+d_{200}+2 q^2\right)\right]e_iM_i^2\nonumber\\
&&\hspace*{1.3cm}
+v^4 \left[d_{012} \left(d_{101}+2 q\right)+m_+ \left(4 q-2 d_{012}\right)+d_{012}^2\right.\nonumber\\
&&\hspace*{2.1cm}\left. \left. -2 \left(d_{022}+q d_{101}+d_{101}^2-d_{200}\right)\right]q_i
\right]f_i,\\
\bcal_{21}&=&-\bcal_{12}^*,\\
\bcal_{22}&=&-\bcal_{11}.
\eea
Here, $f_i$ refers to the $H_iW^+H^-$ coupling \cite{Grzadkowski:2014ada}.
We see that the vanishing of all four matrix elements is achieved by requiring $\bcal_{11}=\bcal_{12}=0$. Since $\bcal_{12}$ has both a real and an imaginary part, the vanishing of $\bcal_{12}$ is equivalent to requiring $|\bcal_{12}|^2=0$. We find
\bea
v^{12}|{\cal B}_{12}|^2&=&
b_{12}^2+b_{13}^2+b_{23}^2,
\eea
where
\bea
b_{ij}&=&v^4 \left[d_{012} \left(d_{101}+2 q\right)+m_+ \left(4 q-2 d_{012}\right)+d_{012}^2\right. \nonumber\\
&&\hspace*{0.8cm}\left.-2 \left(d_{022}+q d_{101}+d_{101}^2-d_{200}\right)\right](e_jq_i-e_iq_j)\nonumber\\
&&+v^2 \left[d_{012} \left(2 q-d_{101}\right)+m_+ \left(4 q-2 d_{012}\right)\right. \nonumber\\
&&\hspace*{1.1cm}\left.+2 \left(-d_{022}+q d_{101}-d_{101}^2+d_{200}+2 q^2\right)\right]e_ie_j(M_i^2-M_j^2)\nonumber\\
&&+2 v^2 \left(d_{012}-2 q\right)(M_i^2q_ie_j-M_j^2q_je_i)
+2  \left(d_{012}-2 q\right)e_ie_j(M_i^4-M_j^4),
\eea
so the vanishing of $\bcal_{12}$ is equivalent to $b_{12}=b_{13}=b_{23}=0$. Requiring the vanishing of the commutator $\bcal$ is then equivalent to requiring $\bcal_{11}=b_{12}=b_{13}=b_{23}=0$. At first glance this looks like four constraints. As we will soon see, these four constraints are not all independent, so the number of independent constraints arising from demanding the vanishing of the commutator $\bcal$ is less than four. It turns out that the detailed analysis becomes easier if we split the analysis in separate parts, depending on whether $\Im J_1$ is vanishing or not\footnote{If we evaluate $\Im J_1$ in the symmetry basis ($\lambda_6=\lambda_7=0$) for the most general form of the VEV, we find
\bea
\Im J_1&=&\frac{v_1^2v_2^2(\lambda_2-\lambda_1)\Im(\lambda_5 e^{2i\xi})}{v^4}.
\eea
The only realistic way that $\Im J_1$ can vanish is if $\Im(\lambda_5 e^{2i\xi})=0$, since the vanishing of $v_1$ or $v_2$ implies $m_{12}^2=0$ (no soft breaking of $Z_2$), and $\lambda_2=\lambda_1$ implies CP2 symmetry of $V_4$.
}.

Let's first assume $\Im J_1\neq0$ (implying CP violation).
We may the solve $\bcal_{11}=0$ for $q$ to get
\bea
q=d_{010}-\half d_{012}-d_{101}-\frac{4\,\Im J_{11}+\Im J_2+2\Im J_{30}}{2\Im J_1}.\label{Z2q}
\eea
Next, we insert the expression for $q$ into the expressions for all three $b_{ij}$ to get
\bea
b_{ij}&=&\frac{e_ie_j(M_j^2-M_i^2)-(e_jq_i-e_iq_j)v^2}{(\Im J_1)^2}
\nonumber\\
&&\times
\left[
2\Im J_1	[2(d_{012}+d_{101}-d_{010})\Im J_1+4\,\Im J_{11}+\Im J_2+2\,\Im J_{30}]M_{H^\pm}^2\right.\label{bij}\nonumber\\
&&\hspace*{0.5cm}
-v^2
\bigl\{
2(d_{010}d_{012}-d_{010}d_{101}-d_{022}+d_{200})(\Im  J_1)^2\nonumber\\
&&\hspace*{1.3cm}+[4(2d_{101}-d_{010})\Im J_{11}+(d_{012}-2d_{010}+3d_{101})\Im J_2\nonumber\\
&&\hspace*{7cm}+2(d_{101}-d_{012})\Im J_{30}]\Im J_1\nonumber\\
&&\hspace*{1.3cm}\left.
+(2\,\Im J_{11}+\Im J_2)(4\,\Im J_{11}+\Im J_2+2\,\Im J_{30})
\bigr\}
\right].
\eea
Demanding that all three $b_{ij}$ should vanish, the only viable solution\footnote{If we require $e_ie_j(M_j^2-M_i^2)-(e_jq_i-e_iq_j)v^2=$ for all three combinations of $i,j$, this will imply $\Im J_1=0$, which we have assumed is not the case.} is given by
\bea
&&2\Im J_1	[2(d_{012}+d_{101}-d_{010})\Im J_1+4\,\Im J_{11}+\Im J_2+2\,\Im J_{30}]M_{H^\pm}^2\nonumber\\
&&=v^2
\bigl\{
2(d_{010}d_{012}-d_{010}d_{101}-d_{022}+d_{200})(\Im  J_1)^2\nonumber\\
&&\hspace*{1cm}
+[4(2d_{101}-d_{010})\Im J_{11}+(d_{012}-2d_{010}+3d_{101})\Im J_2+2(d_{101}-d_{012})\Im J_{30}]\Im J_1\nonumber\\
&&\hspace*{1cm}
+(2\,\Im J_{11}+\Im J_2)(4\,\Im J_{11}+\Im J_2+2\,\Im J_{30})
\bigr\}.\label{Z2Mch}
\eea

We must also consider the possible physical configurations where $\Im J_1=0$, in which case we cannot solve (\ref{B11}) for $q$. We provided a list of all such configurations in Section 3.5 of \cite{Ferreira:2020ana}. We then work systematically through all six possible configurations to see what additional constraints are needed in order to make the commutator $\bcal$ vanish.

For Configuration 1 (the case of full mass degeneracy (two constraints) -- which is CP conserving), we find that the commutator $\bcal$ vanishes whenever $\qcal^2=0$ (this is Case \textoverline{ABBB} from \cite{Ferreira:2020ana}, representing a $U(1)$ invariant (and thereby also $Z_2$ invariant) potential, which we discard since we are only interested in softly broken $Z_2$). Another possibility for the commutator $\bcal$ to vanish in the case of full mass degeneracy is whenever
\bea
2v^4(d_{200}-d_{101}^2)+(v^2d_{101}+M_1^2-2\mchsq)(M_1^2-2v^2q)=0.\label{Z2conf1}
\eea
We then arrive at a model with three constraints in which the non-RGE-stable full mass-deneracy is combined with the vanishing of the commutator $\bcal$. Such a model is also not RGE-stable and will not be discussed further. For completeness it will be listed in Appendix \ref{App:nonRGE} as Case \textoverline{SOFT-Z2-A}.

For Configuration 2 (two constraints - also CP conserving), we find that $\bcal_{11}=0$ and also $b_{ij}=0$. Let us also assume that $e_i\neq0$, so we can solve for $q_j=e_jq_i/e_i$. Then we find that $b_{jk}$ vanish whenever $b_{ik}$ vanish, since they are now proportional to each other. So we get only one additional constraint $b_{ik}=0$ in order to make $\bcal$ vanish in this case. This leaves us with a total of three constraints, but the partial mass-degeneracy is again non-RGE-stable so we will not discuss this case further.

We should also adress what happens for this configuration whenever $e_i=0$ (and we cannot solve for $q_j$). Then we get $e_jq_i=0$. If $q_i=0$, we find that $b_{ik}=0$, so we get only one additional constraint ($b_{jk}=0$) for $\bcal$ to vanish. If on the other hand $e_j=0$, we find (assuming $q_i^2+q_j^2\neq0$ in order to avoid a $Z_2$-symmetric potential) that $b_{ij}$ and $b_{jk}$ vanish simultaneously, so we get only one additional constraint ($b_{jk}=0$) in order to make $\bcal$ vanish. Again, this is non-RGE-stable and will not be discussed further. Within Configuration 2, we end up with cases that has three or sometimes four constraints, all non-RGE-stable. For completeness we list them in Appendix \ref{App:nonRGE} as Case \textoverline{SOFT-Z2-B}.

For Configuration 3 (two constraints - also CP  conserving), we find that we have quartic $Z_2$-invariance if in addition $b_{ij}=0$. This is RGE-stable.

For Configuration 4 (two constraints), we find that demanding $\bcal_{11}=0$, will imply Configuration 1, 2 or 3, already analyzed.

For Configuration 5 (two constraints), we find that demanding $\bcal_{11}=0$, will imply Configuration 2 or 3, already analyzed.

For Configuration 6 (one constraint), we find that we have quartic $Z_2$-invariance if in addition
\bea
v^2(q_je_i-q_ie_j)-e_ie_j(M_i^2-M_j^2)=0,\quad\text{and}\quad q=\half d_{012}.
\eea
This implies CP violation. The total number of constraints is three, and it is listed in Section \ref{sect:softZ2} as Case SOFT-Z2-Y.

\section{Alignment}
\label{sec:soft-alignment}

As we have seen, the symmetry conditions of the quartic potential are complicated,
therefore in this appendix we are going to consider their alignment limit (AL) which,
in our formalism, is simply given by
\bea
e_1=v,\quad e_2=e_3=0\,.
\label{al_def}
\eea
For each symmetry, we shall formulate criteria for having softly broken symmetry in the AL.

In certain cases alignment might be a consequence of a symmetry of the potential, so that (\ref{al_def}) appears naturally without the necessity of parameter tuning. Such possibilities have been discussed within the 2HDM in refs.~\cite{Pilaftsis:2016erj,Darvishi:2020teg,Draper:2020tyq} and generalized for nHDM in \cite{BhupalDev:2014bir,Darvishi:2019ltl}.

\subsection{Softly broken \cp1 in the alignment limit}
\label{sec:cp1-soft-alignment}
The alignment condition (\ref{al_def}) makes $\Im J_1=\Im J_{11}=\Im J_2=0$, but $\Im J_{30}$ may still be non-zero, making room for CP violation in the scalar couplings.
\bea
\left(\Im J_{30}\right)_{\text{AL}}=\frac{q_2q_3(M_3^2-M_2^2)}{v^4},
\eea
implying $q_2q_3(M_3^2-M_2^2)\neq0$ in order for our model to be CP violating in the AL. Next, let us look at the (simplified) expression for $I_{6Z}$ in the AL. First,
\bea
(\Delta m_+)_\text{AL}&=&\frac{vq_1-M_1^2}{2v^2},\\
(\Delta q)_\text{AL}&=&\frac{1}{2}
\left(
\frac{q_2^2}{M_2^2}+\frac{q_3^2}{M_3^3}+\frac{M_1^2}{v^2}
\right).
\eea
We find
\begin{align} \label{Eq:I_6Z_AL}
	(I_{6Z})_\text{AL}&=\frac{\left(\Im\, J_{30}\right)_\text{AL}}{v^4}
	\biggl[
	16v^4(\Delta m_+)_{\text{AL}}^2{(\Delta q)_{\text{AL}}}+8v^4\left(\frac{q_2^2}{M_2^2}+\frac{q_3^2}{M_3^2}\right){(\Delta m_+)}_{\text{AL}}^2 \nonumber\\
	&\hspace*{2.7cm}
	-16v^2(M_2^2+M_3^2){(\Delta m_+)_{\text{AL}}}{(\Delta q)_{\text{AL}}} \nonumber \\
	&\hspace*{2.7cm}-8v^2\left(\frac{q_2^2M_3^2}{M_2^2}+\frac{q_3^2M_2^2}{M_3^2}\right){(\Delta m_+)_{\text{AL}}}
	+16M_2^2M_3^2{(\Delta q)_{\text{AL}}}
	\biggr].
\end{align}
There are different ways to realize $\left(I_{6Z}\right)_{\text{AL}}=0$, which is the condition for invariance of the quartic part of the potential in the AL.
\begin{enumerate}
	\item $\left(I_{6Z}\right)_{\text{AL}}$ vanishes when $\left(\Im J_{30}\right)_{\text{AL}}$ vanish. In this situation we have a CP1-invariant potential as well as a CP-invariant vacuum. There is no CP violation in the scalar sector.
	\item $\left(I_{6Z}\right)_{\text{AL}}$ also vanishes whenever ${(\Delta m_+)}_{\text{AL}}={(\Delta q)}_{\text{AL}}=0$. We know that this corresponds to a CP1-invariant potential. If $\left(\Im J_{30}\right)_{\text{AL}}$ is nonzero, we have spontaneous CP  violation in the AL.
	\item There are other ways to make $\left(I_{6Z}\right)_{\text{AL}}$ vanish than the two mentioned above. If $\left(\Im J_{30}\right)_{\text{AL}} \neq 0$ then the condition $\left(I_{6Z}\right)_{\text{AL}}=0$  becomes a relatively simple quadratic equation\footnote{One can also consider (\ref{Eq:I_6Z_AL}) as a linear equation in $(\Delta q)_\text{AL}$, but the coefficient of front of $(\Delta q)_\text{AL}$ can vanish, so it cannot always be solved for $(\Delta q)_\text{AL}$.} that allows for a determination of ${(\Delta m_+)}_{\text{AL}}$ in terms of model parameters. Therefore the charged Higgs boson mass is given in terms of other parameters. These cases yield models with softly broken CP1 in the AL.
\end{enumerate}

\begin{alignat}{2}
	&\text{\bf Case (SOFT-CP1)$_\text{AL}$:} &\ \
	&  e_1=v,\quad e_2=e_3=0,\quad \left(I_{6Z}\right)_{\text{AL}}=0 \text{ and $q_2q_3(M_3^2-M_2^2)\neq0$}\nonumber\\
	& & & \text{and ${(\Delta m_+)}_{\text{AL}}^2+{(\Delta q)}_{\text{AL}}^2\neq0$.}  \nonumber
\end{alignat}
\subsection{Softly broken $\z2$ in the alignment limit}
\label{sec:Z2-soft-alignment}
Case SOFT-Z2-X requires $\Im J_1\neq0$, and Case SOFT-Z2-Y requires all $e_m\neq0$, none of which are compatible with the AL. Thus, we are left with only Case SOFT-Z2-C, being compatible with alignment,

\begin{alignat}{2}
	&\text{\bf Case (SOFT-Z2-C)$_\text{AL}$:} &\ \
	&  e_1=v,\quad e_2=e_3=q_3=0,\nonumber\\
	& & & (M_1^2-2v^2q)(2\mchsq+M_1^2-2M_2^2-vq_1)=2q_2^2v^2.  \nonumber
\end{alignat}

\subsection{Softly broken \u1 in the alignment limit}
\label{sec:u1-soft-alignment}
When we impose alignment on the softly broken U(1) model, we find
\begin{alignat}{2}
	&\text{\bf Case (SOFT-U1-C)$_\text{AL}$:} &\ \
	&e_1=v,\quad e_2=e_3=q_3=0,\nonumber\\
	& & &2\mchsq =vq_1-M_1^2+2M_3^2
	,\nonumber\\
	& & & 2v^2(M_2^2-M_3^2)q=v^2q_2^2+M_1^2(M_2^2-M_3^2).\nonumber
\end{alignat}

\subsection{Softly broken CP2 in the alignment limit}
\label{sec:cp2-soft-alignment}
When we impose alignment on the softly broken CP2 model, we find
\begin{alignat}{2}
	&\text{\bf Case (SOFT-CP2)$_\text{AL}$:} &\ \
	&e_1=v,\quad e_2=e_3=q_2=q_3=0,\nonumber\\
	& & & 2v^2q=M_1^2.\nonumber \\
	&\text{\bf Case (SOFT-CP2-C)$_\text{AL}$:} &\ \
	&e_1=v,\quad e_2=e_3=q_2=q_3=0,\nonumber\\
	& & & 2v^2q=M_1^2.\nonumber \\
	&\text{\bf Case (SOFT-CP2-D)$_\text{AL}$:} &\ \
	&e_1=v,\quad e_2=e_3=q_2=q_3=0,\nonumber\\
	& & & 2v^2q=M_1^2, \quad 2\mchsq =vq_1.\nonumber \\
	&\text{\bf Case (SOFT-CP2-CC)$_\text{AL}$:} &\ \
	&e_1=v,\quad e_2=e_3=q_2=q_3=0,\nonumber\\
	& & & 2v^2q=M_1^2.\nonumber \\
	&\text{\bf Case (SOFT-CP2-CD)$_\text{AL}$:} &\ \
	&e_1=v,\quad e_2=e_3=q_2=q_3=0,\nonumber\\
	& & & 2v^2q=M_1^2, \quad 2\mchsq =vq_1-M_1^2.\nonumber
\end{alignat}
\subsection{Softly broken CP3 in the alignment limit}
\label{sec:cp3-soft-alignment}
By identifying $\{i,j,k\}=\{1,2,3\}$, we see that Case SOFT-CP3-BCC is already aligned, and by imposing alignment Case SOFT-CP3-B becomes a specialized case of SOFT-CP3-BCC. SOFT-CP3-C simplifies in the AL, so we have two different models of softly broken CP3 in the AL,
\begin{alignat}{2}
	&\text{\bf Case (SOFT-CP3-BCC)$_\text{AL}$:} &\
	&e_1=v,\quad e_2=e_3=q_2=q_3=0,\quad M_2=M_3,\quad  2v^2q=M_1^2.\nonumber\\
	&\text{\bf Case (SOFT-CP3-C)$_\text{AL}$:} &\ \
	&e_1=v,\quad e_2=e_3=q_2=q_3=0,\nonumber\\
	& & & 2v^2q=M_1^2,\quad 2\mchsq=vq_1-M_1^2+2M_3^2.\nonumber \\
	&\text{\bf Case (SOFT-CP3-CC)$_\text{AL}$:} &\ \
	&e_1=v,\quad e_2=e_3=q_2=q_3=0,\nonumber\\
	& & & 2v^2q=M_1^2,\quad 2\mchsq=vq_1-M_1^2+2M_3^2.\nonumber \\
	&\text{\bf Case (SOFT-CP3-B)$_\text{AL}$:} &\ \
	&e_1=v,\quad e_2=e_3=q_2=q_3=0,\quad M_2=M_3,\nonumber\\
	& & & 2v^2q=M_1^2,\quad 2\mchsq=vq_1+M_1^2.\nonumber \\
	&\text{\bf Case (SOFT-CP3-BCC)$_\text{AL}$:} &\ \
	&e_1=v,\quad e_2=e_3=q_2=q_3=0,\quad M_2=M_3,\nonumber\\
	& & & 2v^2q=M_1^2.\nonumber \\
	&\text{\bf Case (SOFT-CP3-CD)$_\text{AL}$:} &\ \
	&e_1=v,\quad e_2=e_3=q_2=q_3=0,\quad M_2=M_3,\nonumber\\
	& & & 2v^2q=M_1^2,\quad 2\mchsq=vq_1-M_1^2.\nonumber \\
	&\text{\bf Case (SOFT-CP3-C$_0$D)$_\text{AL}$:} &\ \
	&e_1=v,\quad e_2=e_3=q_2=q_3=0,\quad M_3=0,\nonumber\\
	& & & 2v^2q=M_1^2,\quad 2\mchsq=vq_1-M_1^2.\nonumber
\end{alignat}
\subsection{Softly broken SO(3) in the alignment limit}
\label{sec:so3-soft-alignment}
By identifying $\{i,j,k\}=\{1,2,3\}$, we see that case SOFT-SO3-BCC is already aligned, and by imposing alignment case SOFT-SO3-A  becomes a specialized case of SOFT-SO3-BCC.
\begin{alignat}{2}
	&\text{\bf Case (SOFT-SO3-BCC)$_\text{AL}$:} &\
	&e_1=v,\quad e_2=e_3=q_2=q_3=0,\quad M_2=M_3,\nonumber\\
	& & &2M_{H^\pm}^2=e_1q_1-M_1^2+2M_2^2,\quad 2v^2q=M_1^2.\nonumber
\end{alignat}

A recent paper \cite{Arco:2022jrt} studies a model with softly broken $\z2$. In particular, the alignment limit is adopted and the impact of the $\z2$-breaking parameter $m_{12}$ on the digamma rate, $h\to\gamma\gamma$ is explored. This corresponds to Case (SOFT-Z2-C)$_\text{AL}$, where $H_3$ decouples from the charged pair ($q_3=0$) in addition to its decoupling (together with $H_2$) from the gauge bosons ($e_2=e_3=0$). In addition, there is a constraint involving $M_1$, $M_2$, $M_{H^\pm}$, $q_1$ and $q_2$, as well as the quartic coupling $q$. The authors of Ref.~\cite{Arco:2022jrt} analyse the digamma rate in terms of the trilinear coupling
\begin{equation}
\lambda_{hH^+H^-}=\frac{1}{v^2}\left[2M^2_{h^\pm}+M_1^2-\frac{2m_{12}^2}{\sin\beta\cos\beta}\right].
\end{equation}
In this model, as we see above, $m_{12}$ has no simple interpretation. Actually, the additional constraint for this model is
\begin{equation}
\frac{q_1}{v}=\frac{1}{v^2}
\left[2M^2_{h^\pm}+M_1^2-2M_2^2-\frac{2v^2q_2^2}{M_1^2-2v^2q}
\right]
\end{equation}
with $\lambda_{hH^+H^-}=q_1/v$. But one should be aware that the mapping of $m_{12}$ to physical parameters depends on the model assumptions (see the other models above).
\section{The non-RGE-stable cases}
\label{App:nonRGE}
\subsection{Softly broken $Z_2$}
\label{App:nonRGEZ2}
\begin{alignat}{2}
	&\text{\bf Case \textoverline{SOFT-Z2-A}:} &\ \
	&\text{$M_1=M_2=M_3$, (\ref{Z2conf1}) apply, and none of the six cases of $\z2$ invariant}\nonumber \\
	& & &	\text{potential found in~\cite{Ferreira:2020ana} apply, for any combination of $i,j,k$.}\nonumber\\
	&\text{\bf Case \textoverline{SOFT-Z2-B}:} &\ \
	& \text{$M_i=M_j$, $e_jq_i-e_iq_j=0$, $b_{ik}=b_{jk}=0$, and none of the six cases of $\z2$}\nonumber \\
	& & &	\text{invariant potential found in~\cite{Ferreira:2020ana} apply, for any combination of $i,j,k$.}\nonumber
\end{alignat}

\subsection{Softly broken U(1)}
\label{App:nonRGEU1}
\begin{alignat}{2}
	&\text{\bf Case \textoverline{SOFT-U1-B}:} &\ \
	&M_i=M_j,\quad
	e_iq_j-e_jq_i=0,\nonumber\\
	& & & 2(e_i^2+e_j^2)M_{H^\pm}^2=(e_i^2+e_j^2)M_i^2+v^2(e_iq_i+e_jq_j),\nonumber\\
	& & &2(e_i^2+e_j^2)(M_i^2-M_k^2)v^2 q=(e_i^2+e_j^2)(M_i^2-M_k^2)M_i^2\nonumber\\
	& & &\hspace*{4.7cm}-v^2\left[(e_kq_i-e_iq_k)^2+(e_kq_j-e_jq_k)^2\right],\nonumber\\
	& & & \text{and none of the four cases of U(1) invariant}\nonumber \\
	& & & \text{potential found in~\cite{Ferreira:2020ana} apply, for any combination of $i,j,k$.}\nonumber
\end{alignat}

\subsection{Softly broken CP3}
\label{App:nonRGECP3}
\begin{alignat}{2}
	&\text{\bf Case \textoverline{SOFT-CP3-ABBB}:} &\ \
	&M_1=M_2=M_3,\quad
	{\qcal}^2=0,\quad 2v^2 q=M_1^2,\nonumber\\
	& & & \text{and none of the five cases of CP3 invariant}\nonumber \\
	& & & \text{potential found in~\cite{Ferreira:2020ana} apply, for any combination of $i,j,k$.}\nonumber
\end{alignat}

\bibliography{biblio}
\bibliographystyle{JHEP}

\end{document}